\documentclass[aps,prx,preprint,superscriptaddress,longbibliography]{revtex4-2}

\usepackage{graphicx}
\usepackage{dcolumn}
\usepackage{bm}
\usepackage{amsmath}
\usepackage{mathrsfs}
\usepackage{stmaryrd}
\usepackage{url}
\usepackage{amssymb}
\usepackage{siunitx}
\usepackage{xcolor}
\usepackage{comment}
\PassOptionsToPackage{hyphens}{url}\usepackage[colorlinks,bookmarks=false,citecolor=blue,linkcolor=blue,urlcolor=blue]{hyperref}
\usepackage{braket}
\usepackage{stackrel}
\usepackage{breqn}
\usepackage{stmaryrd}
\usepackage{wrapfig}
\usepackage{multirow}
\usepackage[normalem]{ulem}
\usepackage[nottoc,numbib]{tocbibind}
\usepackage{caption}
\usepackage{cleveref}
\crefname{extfig}{extended data fig.}{extended data figs.} 
\Crefname{extfig}{Extended data fig.}{Extended data figs.}

\usepackage{adjustbox}
\usepackage{csquotes}
\usepackage{subfiles}
\usepackage[normalem]{ulem} 
\usepackage{color}
\usepackage{lineno}

\DeclareSIUnit\pixel{pixel}

\begin{document}

\title{Measuring non-Abelian quantum geometry and topology in a multi-gap photonic lattice}

\author{Martin Guillot}

\author{Cédric Blanchard}

\author{Martina Morassi}

\author{Aristide Lema\^itre}

\author{Luc Le Gratiet}

\author{Abdelmounaim Harouri}

\author{Isabelle Sagnes}

\affiliation{Center for Nanoscience and Nanotechnology$,$ CNRS$,$ Paris-Saclay university$,$ 91120 Palaiseau$,$ France}

\author{Robert-Jan Slager}

\affiliation{Department of Physics and Astronomy$,$ University of Manchester$,$
Oxford Road$,$ Manchester M13 9PL$,$ United Kingdom}

\affiliation{TCM Group$,$ Cavendish Laboratory$,$ University of Cambridge$,$ JJ Thomson Avenue$,$ Cambridge CB3 0HE$,$ United Kingdom}

\author{F.~Nur \"Unal}

\affiliation{School of Physics and Astronomy$,$ University of Birmingham$,$ Edgbaston$,$ Birmingham B15 2TT$,$ United Kingdom\looseness=-1}

\affiliation{TCM Group$,$ Cavendish Laboratory$,$ University of Cambridge$,$ JJ Thomson Avenue$,$ Cambridge CB3 0HE$,$ United Kingdom}

\author{Jacqueline Bloch}

\author{Sylvain Ravets}
\email[]{sylvain.ravets@c2n.upsaclay.fr}

\affiliation{Center for Nanoscience and Nanotechnology$,$ CNRS$,$ Paris-Saclay university$,$ 91120 Palaiseau$,$ France}


\begin{abstract}
\textbf{Recent discoveries in semi-metallic multi-gap systems featuring band singularities have galvanized enormous interest~\cite{Ahn2019, Wu2019, Bouhon2019, Bouhon2020, Unal2020, Jiang2021, Guo2021, Zhao2022, Peng2022, Yang2024, Slager2024} in particular due to the emergence of non-Abelian braiding properties of band nodes~\cite{Wu2019, Ahn2019, Bouhon2019}. This previously uncharted set of topological phases necessitates novel approaches to probe them in laboratories, a pursuit that intricately relates to evaluating non-Abelian generalizations of the Abelian quantum geometric tensor (QGT) that characterizes geometric responses~\cite{Provost1980, Resta2011, Ma2010, Palumbo2021}. Here, we pioneer the direct measurement of the non-Abelian QGT. We achieve this by implementing a novel orbital-resolved polarimetry technique to probe the full Bloch Hamiltonian of a six-band two-dimensional (2D) synthetic lattice, which grants direct experimental access to non-Abelian charges, the Euler curvature, and the non-Abelian quantum metric associated with all bands. Quantum geometry has been highlighted to play a key role on macroscopic phenomena ranging from superconductivity in flat-bands~\cite{Peotta2015, Torma2022}, to optical responses~\cite{Gao2019, Ahn2020, Bouhon2023, Jankowski2023, Tanaka2024}, transport~\cite{Fang2023, Li2024b, Jankowski2024a, Mandal2024, Mercaldo2025}, metrology~\cite{Yu2024, Li2024a}, and quantum Hall physics~\cite{Rhim2020, Lai2021, Gao2023a}. Therefore, our work unlocks the experimental probing of a wide phenomenology of multi-gap systems, at the confluence of topology, geometry and non-Abelian physics.}
\end{abstract}

\maketitle

\newpage

The last decades have witnessed rapid progress in the universal understanding of topological insulators and metals~\cite{Hasan2010, Chiu2016}. Many of these advances have been linked to two-band models, which provide the minimal framework to describe topological phenomena such as quantum Hall effect. However, recently, characterizations going by the name of multi-gap phases have revealed topologies where multiple gaps get intertwined~\cite{Bouhon2018, Bouhon2019, Wu2019, Ahn2019, Bouhon2020}. Reaching beyond previously known classifications, this gives rise to new topological invariants associated to groups of bands and band singularities, non-Abelian Berry curvatures, and novel physical responses~\cite{Unal2020, Yang2020, Wang2021, Jiang2021, Guo2021, Peng2022, Zhao2022, Slager2024, Breach2024,  Jankowski2024a, Yang2024, Kobayashi2025}.

These theoretical developments have instigated a treasure hunt for new topological characterizations, which are now being related to the underlying topology and geometry of multi-band systems in a broader context, such as generic approaches to defining quantum geometric tensors~\cite{Provost1980, Ma2010, Resta2011, Palumbo2021, Bouhon2023}, multi-gap quantized optical responses~\cite{Jankowski2023, Jankowski2024a}, and multi-gap projective entangled pair states with interacting counterparts~\cite{Wahl2024}. A key exciting agenda is to identify observable signatures in experimentally feasible settings~\cite{Unal2020, Jiang2021, Guo2021,Yang2024, Zhao2022, Peng2022, Hu2024, Breach2024, Slager2024}. Indeed, impressive manifestations of underlying non-Abelian charges or braiding mechanisms have been demonstrated in metamaterials~\cite{Yang2020, Jiang2021, Guo2021, Yang2024} and contingent, yet distinct, signatures observed in quench dynamics of multi-gap Euler insulators~\cite{Zhao2022}.

Probing the topology of a system in the most direct and complete way requires techniques to measure the quantum geometry of the Bloch eigenstates, from which all topological quantities can be computed~\cite{Provost1980, Ma2010,  Resta2011, Palumbo2021, Bouhon2023}. In two-band systems, tremendous advances have been realized using polarimetry techniques to access the eigenstates, culminating with measurements of the (Abelian) quantum geometric tensor~\cite{Li2016, Flaschner2016, Gianfrate2020, Cuerda2024, Kim2025, Guillot2025}. To unlock the experimental exploration of non-Abelian topologies, a key challenge is to develop new methods to access the Bloch eigenmodes in many-band systems. In this work, we experimentally demonstrate the direct probing of the non-Abelian charges and the full non-Abelian band topology of a six-band photonic lattice. Through the use of a novel orbital polarimetry technique that resolves all the multi-component Bloch modes in amplitude and phase, we measure the non-Abelian QGT and explore the intricate physics of band touching points in multi-gap systems.

\section{Multi-gap Euler topology and geometry}

\begin{figure}[t]
    \includegraphics[width=1\linewidth]{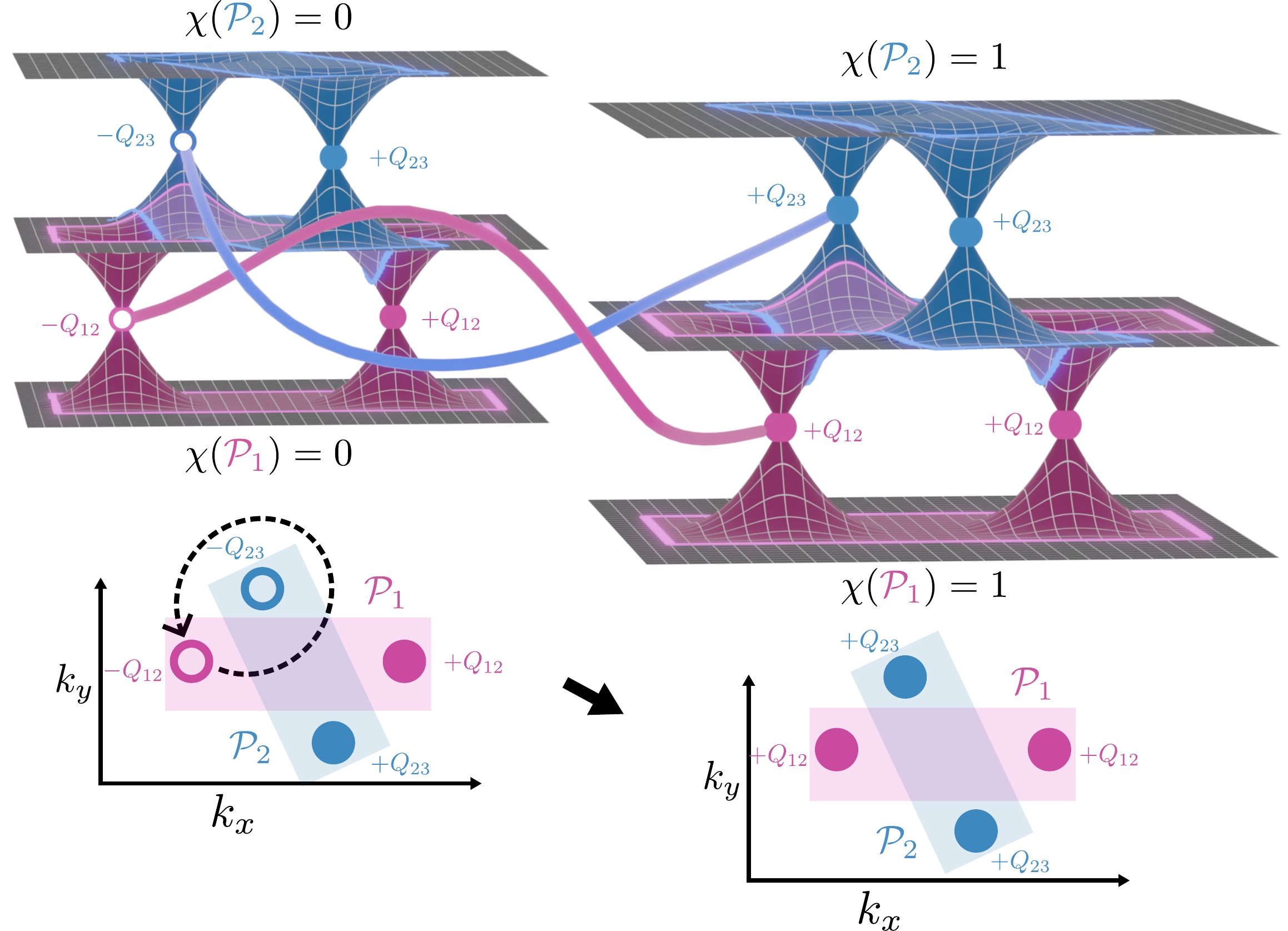}
    \caption{Illustration of momentum space braiding in a multi-gap system (two gaps in between three bands). Band nodes carry non-Abelian charges labeled as $\pm Q_{12},\pm Q_{23}$. In the left panel, nodes come in pairs of opposite charges (empty/filled circles) which can annihilate each other and open a gap. This is captured by vanishing Euler class within the colored momentum patches ($\mathcal{P}_{1,2}$). Moving band nodes in adjacent gaps around each other (red and blue solid lines) produces a modified band structure with similarly charged nodes in each gap (right panel). Such braiding mechanism is illustrated in the momentum space projections in the bottom panels (dashed arrow). After braiding, the nodes cannot annihilate within $\mathcal{P}_{1,2}$, which is captured by a nonzero Euler class $\chi(\mathcal{P}_{1,2})$. Note that for the 3-band system represented here, these charges can be labeled by the quaternion charges $\mathbb{Q}\!=\!\{1,-1,\pm i,\pm j, \pm k\}$.
    }
    \label{Fig:braiding}
\end{figure}

Our focus is on Euler class topology, which emerges in many-band lattices featuring degeneracies in their band structure, as schematically shown in~\cref{Fig:braiding}. In systems with $\mathcal{C}_2 \mathcal{T}$ or $\mathcal{P} \mathcal{T}$ (i.e.~two-fold rotation or parity combined with time-reversal) symmetry, which enforce a reality condition to their Bloch Hamiltonian, band nodes are characterized by non-Abelian frame charges, signifying how the orthonormal frame of eigenvectors rotates around the degeneracies. For 3-band systems these charges can be labeled by the quaternion group $\mathbb{Q}\!=\!\{1,-1,\pm i,\pm j, \pm k\}$~\cite{Wu2019, Bouhon2019}, satisfying the following commutation rules: nodal charges between pairs of nodes located in adjacent bands anti-commute, while they commute otherwise.

For $N>3$ bands the nodes are characterized by generalized quaternions, related to the so-called Salingaros vee group as homotopy characterization of the flag manifold $O(N)/O(1)^N$. More specific to the $N=6$ case, one can employ the anti-commuting ($\{e_i,e_j\}=-2\delta_{i,j}$) Clifford algebra elements of $Cl_{0,5}=\{e_1,e_2,e_3,e_4,e_5\}$, and define generalized quaternion charges as $Q_{12}=e_1$ between band 1 and 2, and $Q_{n,n+1}=e_{n-1}e_{n}$ between bands $n$ and $n+1$ with $n=\{2,3,4,5\}$. One also has charges spanning across multiple band gaps formed as products (double or triple) of any $Q_{n,n+1}$, which altogether result in 34 conjugacy classes ($\{+1\}, \, \{-1\}, \, \bigcup_{1\leq n<n'\leq 6}\{Q_{n,n'}\}$, \, $\bigcup_{1\leq n<n'<n''<n'''\leq 6}\{Q_{n,n',n'',n'''}\}, \, \{Q_{123456}\}, \, \{-Q_{123456}\}$) and recover the desired commutation relations~\cite{Wu2019}. Geometrically, $Q_{n,n+1}$ relates to a $\pi$ rotation of the orthonormal frame of eigenvectors due to a band node between bands $n$ and $n+1$ as will be probed in our work below.

We illustrate these non-Abelian charges in~\cref{Fig:braiding} with circles in distinct colors representing different `gaps' (two-band subspaces) and empty (filled) markers representing the negative (positive) non-Abelian charge, e.g.~$-Q_{n,n+1}~(+Q_{n,n+1})$. The left panel in~\cref{Fig:braiding} shows a situation where nodes are created from vacuum in pairs with opposite charges. Upon deformation of the lattice, these nodes can be merged and annihilate each other, leading to the opening of a gap~\cite{Wunsch2008, Montambaux2009, Montambaux2018, Ahn2019, Bouhon2019}. One remarkable feature of multi-gap systems stems from the fact that these non-Abelian charges can be changed by braiding the nodes in adjacent gaps around each other in momentum space~\cite{Bouhon2019}, which results in pairs with identical charge in a given gap (see right panel in~\cref{Fig:braiding}). These nodes cannot annihilate each other upon combination. This obstruction against annihilation within a momentum patch $\mathcal{P}$ is precisely captured by a non-zero value of the quantized Euler class $\chi_{n,n+1}\in \mathbb{Z}$ defined as:
\begin{equation}
\label{eq:Eulerpatch}
\chi_{n,n+1} (\mathcal{P}) = \frac{1}{2\pi} \!\left[\int_{\mathcal{P}} \! \mathrm{Eu}_{n,n+1}(\bm{k}) \,dk^2 \!-\! \oint_{\partial \mathcal{P}} \! \mathcal{A}_{n,n+1}(\bm{k}) \!\cdot\! d\bm{k} \right]\!,
\end{equation}
where $\bm{k}$ is the momentum. We here employ the Euler connection one-form $\mathcal{A}_{n,n+1}^i \!=\!\langle u_{n,\bm{k}} \vert \partial_{k_i} u_{n+1,\bm{k}} \rangle$, and the non-Abelian Berry curvature $\mathrm{Eu}_{n,n+1}  \!=\! \left\langle \partial_{k_x} u_{n,\bm{k}} | \partial_{k_y} u_{n+1,\bm{k}}\right\rangle \!-\! \left\langle \partial_{k_y} u_{n,\bm{k}} | \partial_{k_x} u_{n+1,\bm{k}} \right\rangle$ for the Bloch eigenstates $|u_{n,\bm{k}}\rangle$ of band $n$. Both quantities are related through the Pfaffian $\mathrm{Pf}$,
$\mathrm{Eu}_{n,n+1} (\bm{k}) = d \mathrm{Pf} \mathcal{A}_{n,n+1} (\bm{k})$.

As generalization of the Abelian (single-band) metric~\cite{Provost1980, Resta2011}, an $N$-band non-Abelian quantum geometric tensor (QGT)~\cite{Ma2010, Palumbo2021,Bouhon2023} can be defined as $G^{\alpha,\beta}_{i,j}(\bm{k})=\langle \partial_i u_{\alpha,\bm{k}}|(1-\hat{P})| \partial_j u_{\beta,\bm{k}}\rangle$, in terms of the projector onto the $N$ involved bands, $\hat{P}=\sum_{n=1}^{N}|u_{n,\bm{k}}\rangle\langle u_{n,\bm{k}}|$. Specifying the band labels $(\alpha,\beta)$ for the two bands $|u_{n,\bm{k}}\rangle,|u_{n+1,\bm{k}}\rangle$, and $(i,j)\in(k_x,k_y)$, one readily verifies that the imaginary part of $G$ is zero due to the reality condition and that the off-diagonal parts can be combined to form the Euler curvature in \cref{eq:Eulerpatch}. Furthermore, the real symmetric part of QGT is known as the quantum metric $g^{\alpha,\beta}_{i,j}$. Multi-gap topological ideas naturally extend to the broader quantum geometric viewpoints~\cite{Ma2010,Palumbo2021}, where we note that such QGTs even generalize to systems with degenerate bands by using the so-called Pl\"ucker embeddings~\cite{Bouhon2023}.

\section{Experimental reconstruction of the Bloch Hamiltonian}

Our experiments are conducted on lattices of semiconductor optical resonators~\cite{Schneider2017} arranged into a honeycomb pattern containing two sites per unit cell named $A$ and $B$ (see Methods for more details). The building block of these lattices are micron-sized micropillar cavities that support discrete optical modes. The lowest energy mode presents a single radially symmetric lobe ($\ket{s}$ orbital), and the next two degenerate modes exhibit two out-of-phase lobes ($\ket{p_x}$ and $\ket{p_y}$ modes). Within this work, we focus exclusively on these three modes and do not consider higher-energy modes, although extensions are naturally possible.

\begin{figure*}[t]
    \includegraphics[width=\textwidth]{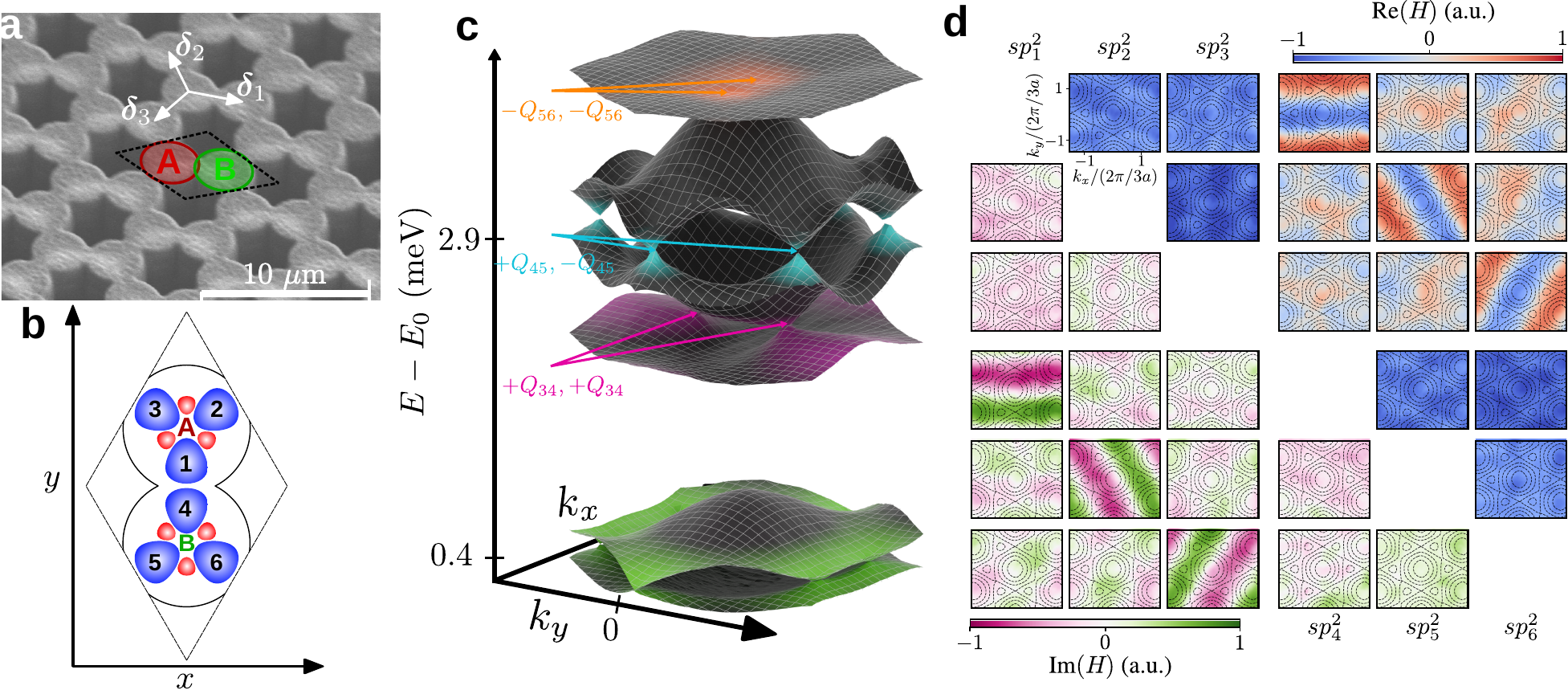}
    \caption{
    \textbf{a}.~Scanning electron microscope image of the honeycomb lattice of coupled micropillars probed in this work. We schematically represent a unit cell (black diamond shape) containing $A$ and $B$ sites. The vectors $\bm{\delta}_i$ connect an $A$ site to its three $B$ neighboring sites.
    \textbf{b}.~Representation of the lattice unit cell showing the $\ket{sp^2}$ modes with positive (negative) lobes in blue (red) color.
    \textbf{c}.~Measured band dispersion, obtained by applying our eigenstate reconstruction method to determine the band energies with sub-linewidth precision. Arrows point toward the different nodes in the $p$-bands. We also indicate their measured generalized quaternion charges.
    \textbf{d}.~Reconstruction of the $\bm{k}$-dependent off-diagonal matrix elements of the Hamiltonian in the $\left\{ \ket{sp^2_{\sigma}} \right\}$ basis. Each plot shows the real (upper triangle) and imaginary (lower triangle) part of one of the Hamiltonian components, as a function of $k_x$ and $k_y$. In each panel, dashed lines show iso-energy contours of the difference between bands 5 and 4.
}
    \label{fig:Graphene}
\end{figure*}

In the honeycomb lattice we study (see microscope image in \cref{fig:Graphene}a), the micropillars are designed with a diameter large enough to resolve the on-site individual orbital profiles. The system is well described by a $6 \times 6$ tight-binding Hamiltonian written in the $\mathcal{B} = \{ \ket{s}, \ket{p_x}, \ket{p_y} \} \otimes \{ \ket{A}, \ket{B} \}$ basis as:
\begin{equation}
    \hat{H}(\bm{k})=
    \begin{bmatrix}
        \hat{H}_s(\bm{k}) & \hat{H}_{sp}(\bm{k}) \\
        \hat{H}_{sp}(\bm{k})^\dagger & \hat{H}_{p}(\bm{k})
    \end{bmatrix}
    \, ,
    \label{eq:6Hamiltonian}
\end{equation}
where $\hat{H}_{s(p)}({\bm{k}})$ is a $2 \times 2 \,(4 \times 4)$ Hamiltonian describing the $s(p)$-bands, and $\hat{H}_{sp}({\bm{k}})$ is the matrix coupling between the $s$ and $p$ sectors (see Methods Appendix~B for full expressions). Diagonalizing this Hamiltonian yields six energy bands and the corresponding six-component Bloch eigenvectors with complex amplitudes $u_{n,\bm{k}}^{\sigma}$ for $\sigma \in \llbracket 1;6 \rrbracket$ in the $\mathcal{B}$ basis.

To implement orbital polarimetry of this six-band system, the key experimental challenge is to probe both the amplitude and the phase of all $u_{n,\bm{k}} ^{\sigma}$ coefficients. To this end, the $\mathcal{B}$ basis is not well suited as its orbitals exhibit strong spatial overlap (see~\cref{fig:PillarOrbitals}a). Inspired by the Linear Combination of Atomic Orbitals, we here introduce a basis of six hybrid orbitals $\ket{sp^2_{\sigma}}$ (see mode profiles in~\cref{fig:PillarOrbitals}a):
\begin{equation*}
    \ket{sp^2_{\sigma}} =  \frac{1}{\sqrt{3}}\left(\ket{s} + \sqrt{2}\left(\sin(\theta_{\sigma})\ket{p_x}+\cos(\theta_{\sigma})\ket{p_y}\right)\right) \, .
    \label{eq:sporbitals}
\end{equation*}
In this basis, each $\ket{sp^2_{\sigma}}$ orbital  presents a main lobe pointing in the direction of a lattice bond (blue lobes in \cref{fig:Graphene}b), and forming an angle $\theta_\sigma$ with respect to the vertical direction. These lobes are well separated within a lattice site with minimal overlap, allowing each orbital to be probed individually. In the rest of the article, we measure the complex amplitudes $v_{n,\bm{k}} ^{\sigma}$ of the Bloch eigenvectors $\ket{v_{n,\bm{k}}}$ written in the $\left\{ \ket{sp^2_{\sigma}} \right\}$ basis.

We now briefly describe the experimental techniques employed to realize the full tomography of the Bloch eigenvectors (see Methods for detailed explanations). We optically excite the sample maintained at cryogenic temperature ($4 \, \unit{K})$ using a non-resonant laser. We collect the photoluminescence signal and project the lattice plane onto a spatial light modulator (SLM) placed between crossed polarizers. The SLM imprints a phase pattern onto the signal, which enables us to selectively modulate the amplitude and phase of the light emitted within different sub-areas of each lattice site. In practice, we divide the unit cell into six circular sectors (see colored circles in \cref{fig:PillarOrbitals}b) and multiply the signal amplitude within each sector by a complex coefficient $m^\sigma$. The light is then directed to a Fourier imaging system, providing energy- and momentum-resolved photoluminescence intensity maps $I_{\bm m} (E,{\bm k})$ under various phase configurations of the SLM, where ${\bm m}$ is the six-component vector gathering all $m^\sigma$.

We show, in the Methods Appendix~D that:
\begin{equation}
     I_{\bm m} (E,{\bm k}) = \sum_{n=1}^6 \eta^n_{\bm{k}}(E)
     \left | \sum_{\sigma}
     m^{\sigma} \, e^{-i \bm{k} \cdot \bm{d}^{\sigma}} \, v^{\sigma}_{n,\bm{k}} \right | ^ 2 \, ,
     \label{eq:IntensitySpectra2}
\end{equation}
where $\eta^n_{\bm{k}}(E)$ is the spectral lineshape in band $n$ at wavevector $\bm{k}$, and $\bm{d}^\sigma$ is the vector of length $|\bm{d}^\sigma| = ({2-\sqrt{3}})a$ connecting the center of a lattice site to the center of the sector $\sigma$ selected by the SLM within that site, as shown in \cref{fig:PillarOrbitals}b. As we detail below, measuring $I_{\bm m} (E,{\bm k})$ for different values of ${\bm m}$ enables us to measure all coefficients of $\ket{v_{n,{\bm k}}}$. This approach is at the core of the recently invented sublattice polarimeter for two-band Hamiltonians~\cite{Guillot2025}.

Here we demonstrate an orbital polarimeter that applies to lattices with multiple orbitals per site. We access the six-component Bloch eigenmodes $\ket{v_{n,\bm{k}}}$ by performing thirty-six $I_{\bm m}(E,{\bm k})$ measurements using properly-chosen configurations of $\bm{m}$. Precisely, we perform (i) six measurements in which only one orbital $\sigma$ contributes, by setting all other elements of $\bm{m}$ to zero ($m^{\sigma'} = \delta_{\sigma,\sigma'}$), (ii) conduct thirty complementary measurements, by using all possible combinations of ${\bm{m}^{\sigma}+\bm{m}^{\sigma'}}$ and $({\bm{m}^{\sigma}+ i \bm{m}^{\sigma'}})/\sqrt{2}$ with $\sigma < \sigma'$ (see \cref{fig:Masks} for exemplary $I_{\bm m}(E,{\bm k})$ measurements).

Having realized the thirty-six measurements, we determine all quadratic quantities of the form $\sum_{n=1}^6 \eta^n_{\bm{k}}(E) v^{\sigma}_{n,\bm{k}} {v^{\sigma'}_{n,\bm{k}}}^*$ (see Methods Appendix~E). These are matrix elements of the operator:
\begin{equation}
     \hat{\rho} (E,{\bm k}) = \sum_{n=1}^6 \eta^n_{\bm{k}}(E) \ket{v_{n,{\bm k}}}\!\!\bra{v_{n,{\bm k}}} \, ,
     \label{eq:IntensitySpectra3}
\end{equation}
expressed in the $\{ sp^2_{\sigma} \}$ basis. This operator is the density matrix describing, for each $E$ and $\bm k$, the (mixed) state of light resulting from the incoherent superposition of emissions from all six bands. For each value of ${\bm k}$, we jointly diagonalize~\cite{Shi2011} all matrices $\hat{\rho} (E, {\bm k})$ for which the intensity exceeds a certain threshold (see Methods), and retrieve the Bloch eigenstates $\ket{v_{n,\bm k}}$. The associated eigenvalues directly yield the intensity profiles $\eta_{\bm k}^n(E)$ from which we extract the energy of each Bloch mode (see \cref{fig:ResolvedSpectra}). 

The resulting measured band dispersion is shown in \cref{fig:Graphene}c and reveals six bands: two lowest-energy $s$-bands, and four higher-energy $p$-bands. We label the degeneracies within adjacent gaps by generalized quaternion charges, with their relative signs experimentally determined as will be explained in the next section. We note that the positions of the nodes reveal a breaking of the lattice $C_6$ symmetry and hint towards the presence of slight anisotropies in the microstructure. As shown in~\cref{fig:3Ddispersions}, the observed dispersion is well reproduced by diagonalizing the tight-binding Hamiltonian including a slight anisotropy in the inter-site couplings ($\beta/t_p = 1.06$) and a small onsite energy splitting ($\epsilon_{\rm el}/t_p = 0.07$) between the $p_x$ and $p_y$ orbitals (see Appendix~C in the Methods section). Moreover, having access to both energies and eigenvectors, we reconstruct the experimental Bloch Hamiltonian, shown in the $\{ sp^2_{\sigma} \}$ basis in \cref{fig:Graphene}d. In the measured Hamiltonian, we identify the couplings between orbitals in the same site or in adjacent sites, reflecting the lattice symmetries. We observe excellent agreement with the tight-binding Hamiltonian shown in~\cref{fig:TBspHamiltonian}. We emphasize that our method provides sublinewidth resolution in the determination of the Bloch eigenmodes and energies, including near the band touching points. This is a crucial asset to investigate the non-Abelian topology of the multi-band lattice.

\section{Direct measurement of the non-Abelian Berry curvature, quantized Euler class and generalized quaternion charges}

\begin{figure*}[t]
    \includegraphics[width=\textwidth]{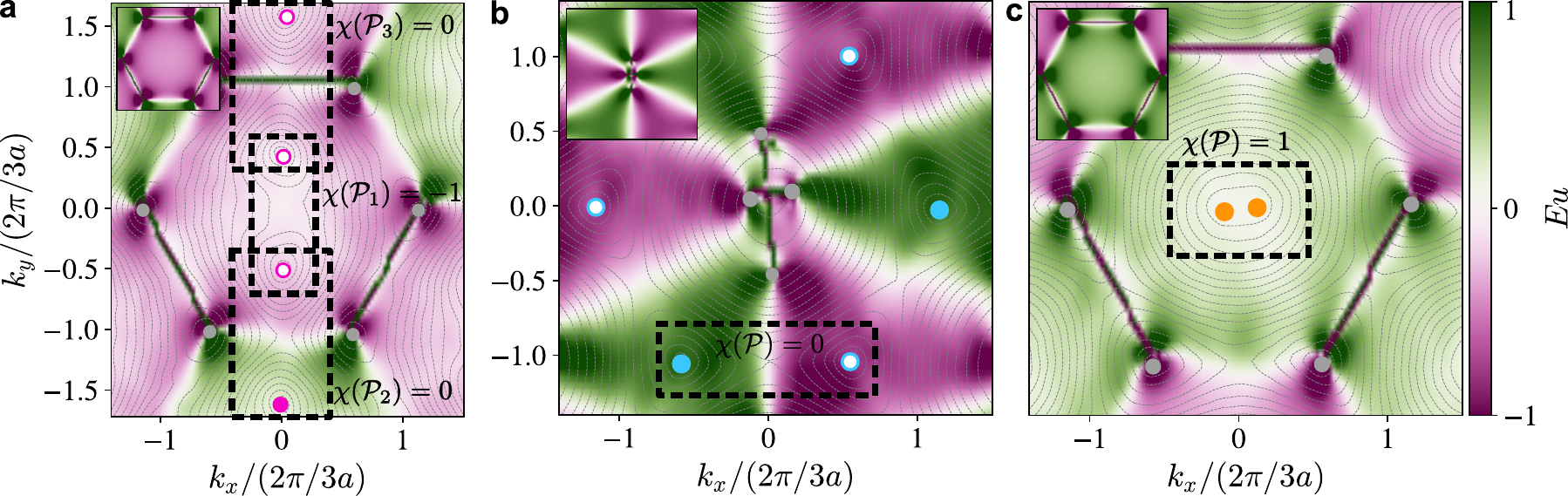}
    \caption{Experimentally reconstructed Euler curvature ${\rm Eu}_{n,n+1} ({\bm k})$ for the pair of bands $p$-bands \textbf{a}.~(3,4), \textbf{b}.~(4,5), and \textbf{c}.~(5,6). In each panel, nodes in adjacent gaps are represented with gray full circles. The Euler class calculated over different patches marked by dashed lines are indicated. Non-zero values of $\chi(\mathcal{P})$ in \textbf{(a,c)} indicates non-Abelian charges with the same sign within these gaps, $-Q_{3,4}$ and $+Q_{5,6}$ respectively. According to the measured Euler class we represent positive (negative) charges in each gap by full (empty) circles. Insets show the tight-binding simulations of the Euler curvature using the following parameters given in units of $t_p$: $\epsilon_s = 0,\: \epsilon_{p} = 5,\: \epsilon_{\rm el} = 0.07,\: t_s = 0.2,\: t_p = 1,\: t_{sp} = 0.2,\: \beta = 1.06$.
    }
    \label{fig:Euler}
\end{figure*}

Having measured all eigenstates and energies of the system, we are now able to experimentally unveil the non-Abelian topology in our multi-gap lattice. Owing to the ${\mathcal C}_2\mathcal{T}$ symmetry, we can adopt a gauge in which the Hamiltonian is real (see Methods Appendix~F). This allows us to use the corresponding real eigenvectors to extract the full non-Abelian QGT, namely the Euler curvature ${\rm Eu}_{n, n+1} ({\bm k})$ (\cref{fig:Euler}) as well as all components of the non-Abelian quantum metric $g^{\alpha,\beta}_{i,j} ({\bm k})$ (\cref{fig:QGT_1}). \Cref{fig:Euler} shows the measured ${\rm Eu}_{n, n+1} ({\bm k})$ for all $p$-band nodes identified in \cref{fig:Graphene}c. Notably, we observe discontinuity lines, called Dirac strings (DSs), connecting adjacent nodes. While their exact location depends on the chosen gauge, DSs arise from the $\pi$-flux of a Dirac cone and yields a $\pi$ shift in the Zak phase along non-contractable paths~\cite{Ahn2019, Bouhon2019}. For clarity, we choose, in each panel of~\cref{fig:Euler}, the gauge that minimizes the size of the DSs.

We first focus on the pair of principal bands $(4,5)$, which hosts band nodes at the $K$ and $K'$ points (blue circles in~\cref{fig:Euler}b). We observe that the measured Euler curvature ${\rm Eu}_{4,5} ({\bm k})$ in this gauge shows alternance of positive and negative values throughout the ${\bm k}$-space. Near the center of the BZ, the grey circles mark the adjacent nodes between bands $(3,4)$ and $(5,6)$, where the Euler curvature is ill-defined. Notably, the adjacent nodes also give rise to highly peaked values of the metric shown in \cref{fig:QGT_1}a-c. Calculating the Euler class thus requires choosing a patch in a region where the two-band subspace is separated from other bands~\cite{Bouhon2019, Jiang2021, Slager2024}. Upon integrating within patch $\mathcal{P}$ (dashed black line) \textit{via}~\cref{eq:Eulerpatch} we find that $\chi_{4,5}(\mathcal{P}) = 0.0 \pm 0.09$. The vanishing of $\chi_{4,5}(\mathcal{P})$ shows that the principle nodes at $K$ and $K'$ carry opposite non-Abelian charges ($\pm Q_{4,5}$ in \cref{fig:Graphene}c, and empty/full blue circles in \cref{fig:Euler}b). As a result, if the nodes were to be moved and merged within patch $\mathcal{P}$, they would annihilate and a gap would open. This can be achieved for example \textit{via} lattice deformation through uniaxial strain~\cite{Wunsch2008, Montambaux2009, Tarruell2012, Bellec2013, Montambaux2018, Peng2022}.

We now focus on the pair of principle bands $(3,4)$, which hosts two band touching points near the center of the Brillouin zone (pink circles in~\cref{fig:Euler}a). These are surrounded by adjacent nodes $(K,K')$ between bands $4$ and $5$ (gray circles in \cref{fig:Euler}a), accompanied by highly peaked value of the metric in \cref{fig:QGT_1}d-f. Upon integrating within patch $\mathcal{P}_1$ \textit{via}~\cref{eq:Eulerpatch}, we find a non-zero integer value of the Euler class $\chi_{3,4}(\mathcal{P}_1) = -1.0 \pm 0.09$. This shows that the two nodes carry the same non-Abelian charges ($-Q_{3,4}$ in \cref{fig:Graphene}c, empty circles in \cref{fig:Euler}a) and are obstructed to annihilate over the patch $\mathcal{P}_1$~\cite{Ahn2019, Bouhon2019, Jiang2021}. If the nodes were to be brought together in momentum space within $\mathcal{P}_1$, e.g.~by applying strain, a parabolic band touching point (type III Dirac cone) will form, as observed experimentally in Ref.~\cite{Milicevic2019}. Notably, the Euler class is patch dependent and vanishes over the $\mathcal{P}_{2,3}$ patches extending over two BZs ($\chi_{3,4}(\mathcal{P}_{2,3}) = 0.0 \pm 0.09$)~\cite{Slager2024}. In the chosen gauge, this vanishing of the Euler class is manifested by either a sign change in ${\rm Eu}_{3,4} ({\bm k})$ or discontinuities in the eigenstates revealed by the presence of a DS within the patch. Finally, bands $(5,6)$ show a behavior similar to that of bands $(3,4)$, also revealing a non-trivial Euler class $\chi_{5,6}(\mathcal{P}) = 1.0 \pm 0.09$ within the first BZ (see \cref{fig:Euler}c), hence, same charges ($+Q_{5,6}$). All data feature very good agreement with the theory prediction for tight-binding calculations of the Euler curvature (see insets in \cref{fig:Euler} and in \cref{fig:QGT_1}).

We emphasize that the observed patch-dependent behavior is a direct consequence of the underlying braiding properties of band nodes~\cite{Bouhon2019}. Indeed, it shows that while the nodes nearby $\Gamma$ are protected within $\mathcal{P}_1$, they can annihilate each other across the BZ. Namely, if the nodes were to be moved and merged~\cite{Montambaux2018} e.g.~within $\mathcal{P}_{3}$ in \cref{fig:Euler}a, this would require crossing the DS in the adjacent gap that flips the sign of the charge~\cite{Ahn2019, Bouhon2019}. As such, DSs offer an alternative way of describing braiding~\cite{Slager2024}, in this case around a $K/K'$ node in the adjacent gap. Notably, while we do not perform braiding operations in this experiment, we effectively probe this physics by measuring the patch-dependent Euler class.

\section{Phase winding and Quantum geometry in reduced sub-spaces} 

Non-abelian charges, as measured in~\cref{fig:Euler}, have been shown to act as $\pm\pi$-fluxes within a given two-band eigenstate sub-space~\cite{Montambaux2018, Breach2024}. Experimentally measuring the full six-band Hamiltonian enables us to gain further insight into the low-energy physics near band-touching points. In particular, we can focus on the vicinity of a given node between the bands $(n,n+1)$ by projecting the measured Hamiltonian onto the relevant 2D subspace spanned by the two bands~\cite{Montambaux2018}. Expanding around a reference point of wavevector ${\bm k_0}$, one can obtain a $2\times2$ effective Hamiltonian acting on the spinor $\{ \ket{u_{n,{\bm k_0} }} , \ket{u_{n+1,{\bm k_0}}}\}$ as:
\begin{equation}
     \hat{H}_{\rm eff}(\bm{q})=\lambda_0(\bm{q}) \hat{\sigma}_0+\bm{\lambda}(\bm{q}) \bm{\hat{\sigma}} \, ,
     \label{eq:EffectiveHamiltonian}
\end{equation}
where ${\bm q} = {\bm k} - {\bm k_0}$, $\hat{\sigma}_0$ is the identity matrix, $\bm{\hat{\sigma}} = \left(\hat{\sigma}_1, \hat{\sigma}_2, \hat{\sigma}_3 \right)$ are Pauli matrices defined relative to the eigenstates at ${\bm k_0}$ (see Methods Appendix~G). The vector $\bm{\lambda}(\bm{q})$ can be viewed as a pseudo-magnetic field on the Bloch sphere with north and south poles representing the two spinor components $ \ket{u_{n,{\bm k_0} }}$ and $\ket{u_{n+1,{\bm k_0}}}$. When varying $\bm{q}$, $\bm{\lambda}(\bm{q})$ evolves rotating around a preferred direction that depends on ${\bm k_0}$~\cite{Montambaux2018}, and that defines the orbital plane along which we follow the azimuthal angle $\varphi_{n,n+1}$ (see \cref{fig:BlochSphere}). Following the evolution of $\varphi_{n,n+1}$, one can measure some of the topological features, such as the phase windings and topological charges~\cite{Montambaux2018, Milicevic2019}.

\begin{figure*}[t]
    \includegraphics[width=\textwidth]{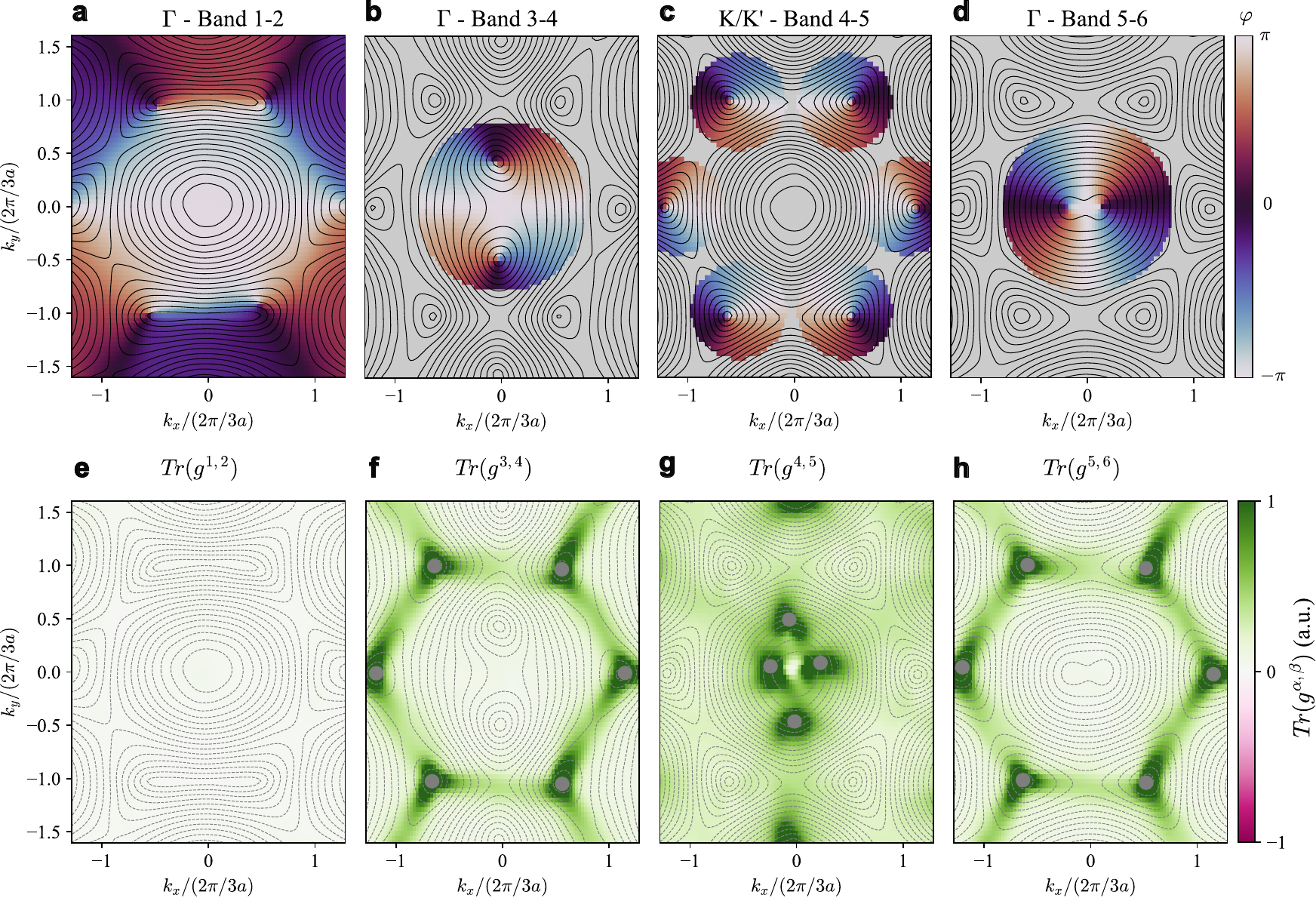}
    \caption{
    \textbf{a-d.}~Effective two-band Hamiltonians around selected ${\bm k_0}$ points (indicated above each panel). The color map represents the azimuthal angle $\varphi_{n,n+1}$ of the vector $\bm{\lambda}(\bm{q})$ in the orbital plane defined in the text (see~\cref{fig:BlochSphere}), plotted over the $\boldsymbol{k}$-space regions where the two-band approximation holds.
    \textbf{e-h.}~Trace of the non-Abelian quantum metric as function of $k_x$ and $k_y$ for the corresponding subsets of bands. In all panels, dashed lines show iso-energy contours of the energy difference between the two considered bands.
    }
    \label{fig:Heff}
\end{figure*}

\Cref{fig:Heff}a-d show in color scale the measured $\varphi_{n,n+1}({\bm k})$ for different pairs of bands, in regions where we have checked that the two-band approximation remains accurate and mode mixing with other bands can be neglected. The validity of the two-band approximation can be further related to the full non-Abelian QGT, since in regions where mode mixing is insignificant, the trace of the quantum metric ${\rm Tr} (g^{n,n+1})$ is expected to be negligible as well. We check this by contrasting the regions selected in \cref{fig:Heff}a-d with the quantum metric trace plotted in~\cref{fig:Heff}e-h. Considering $s$ bands in \cref{fig:Heff}a, we plot the measured $\varphi_{12} ({\bm k})$ obtained by expanding around ${\bm k_0}=\Gamma$. Since the two $s$-bands are well isolated from all other bands, we perform the expansion throughout the entire Brillouin zone, and recover the characteristic phase pattern of graphene bands with opposite $\pm \pi$ phase windings around $K$ and $K'$ points.

We now turn to the band-touching points within the $p$-band manifold, where multi-band nature gets pronounced. Focusing on the nodes between bands $(3,4)$, we construct an effective two-band Hamiltonian by expanding around the $\Gamma$ point. The map $\varphi_{34}({\bm k})$ in \cref{fig:Heff}b, reveals the presence of a pair of band nodes in vicinity of $\Gamma$. Both nodes display a $\pi$ phase winding with the same sign~\cite{Milicevic2019}, in stark contrast to the opposite charges observed within the $s$-bands. This is due to the fact that the two band touching points carry the same non-Abelian charge (see~\cref{fig:Euler}a), which project into the two-band subspace as identical fluxes $\pi$~\cite{Breach2024}. Contrary to what was observed for the $s$-bands, ${\rm Tr}(g^{3,4})({\bm k})$ takes highly-peaked values close to the $K$ points, signaling that the mode mixing in these areas becomes large and limiting the validity range of the two-band expansion. 

Similar behavior is found for the upper pair of $p$-bands $(4,5)$ and $(5,6)$. In \cref{fig:Heff}c, we evaluate $\hat{H}_{\rm eff}$ around each $K$ and $K'$ points taken individually, while keeping the gauge fixed between them. We observe that $\varphi_{45}({\bm k})$ shows opposite windings $\pm \pi$ between the $K$ and $K'$ points, as expected from the measurement of opposite non-Abelian charges in~\cref{fig:Euler}b. Finally, we verify that $\varphi_{56}({\bm k})$ shows two $\pi$ windings of the same sign around the $\Gamma$ point.

\section{Discussion and outlook}

The direct measurements of the geometric and topological structure of eigenstates in multi-band systems illustrate how the relative arrangement of principal and adjacent band touching points influences the Euler class, either preventing or allowing their annihilation. This highlights the intricate nature of multi-band topology involving singularities that are distant both in energy and momentum, and underpins the Euler class as a fundamental tool to characterize the topological protection of band nodes against annihilation within momentum space. A promising research direction within our platform is to gain control over the motion of such band touching points and to braid them by continuous tuning of the lattice parameters~\cite{Bouhon2019, Yang2020, Peng2022, Slager2024, Hu2024, mondal2024b}. In~\cref{fig:ExtdBraiding}, we propose a protocol implementing a scenario analogous to the one depicted in~\cref{Fig:braiding}. We start with a large value of the ellipticity $\epsilon_{\rm el} / |t_p| =3.1$, where the calculated band structure is fully gapped. By continuously tuning down $\epsilon_{\rm el} / |t_p|$, pairs of nodes of opposite charge successively appear within the different gaps. Some charges are eventually made to traverse a DS in the adjacent gap, hence altering their charge. At the end of the braiding protocol ($\epsilon_{\rm el} / |t_p| = 0$) the charge configuration ends up being identical to the one observed in this work, thus highlighting the braiding experienced by the charges that we measure when they are created from vacuum. In Supplementary Movie 1, we illustrate the full protocol, thus outlining a feasible route to implement braiding in future experiments.

This work opens a new experimental playground for further exploring multi-band topology, and the intricate physics of band-touching points. We emphasize that while this work has focused on the case of an orbital lattice, the method is broadly applicable to a wide range of lattice systems featuring multiple sites per unit cell or multiple spinor degrees of freedom, which includes lattices hosting flat bands~\cite{Finck2025}, trigonal warping~\cite{Nalitov2015}, or Chern bands~\cite{Klembt2018}. Moreover, by harnessing the unique tunability and driven-dissipative nature of polariton platforms, our work lays the foundation for future studies of non-Hermitian and nonlinear multi-band topological phenomena~\cite{Wang2021,Konig2023}. Finally, we point out rapid advances in the design and control of Moir\'e materials are bringing within reach effective models, and probably realizable materials, that in the future could host non-Abelian or multi-gap topologies~\cite{Ahn2019,mondal2024b}.


%


\section*{Methods}

\subsection{Sample description}

The sample consists of an epitaxially grown semiconductor microcavity heterostructure. It is composed of two distributed Bragg reflectors (DBRs) made of alternating $\mathrm{Al}_{0.95}\mathrm{Ga}_{0.05}\mathrm{As}$ and $\mathrm{Al}_{0.1}\mathrm{Ga}_{0.9}\mathrm{As}$ $\lambda / 4$ layers, with 26 pairs in the top DBR and 30 in the bottom one. The design wavelength, $\lambda \simeq \SI{860}{\nano\meter}$, sets the center of the DBR high-reflectivity stop band. The DBRs enclose a $\mathrm{GaAs}$ cavity spacer containing a $\SI{17}{\nano\meter}$ $\mathrm{In}_{0.05}\mathrm{Ga}_{0.95}\mathrm{As}$ quantum well positioned at its center. At a temperature of $\SI{4}{\kelvin}$, the exciton resonance energy is $\SI{1450}{\milli\electronvolt}$, approximately $\SI{10}{\milli\electronvolt}$ above the zero-momentum cavity mode energy. Exciton–polaritons are normal modes emerging from the radiative coupling between the excitonic and photonic modes, with Rabi coupling $\Omega = \SI{3.6}{\milli\electronvolt}$.

The honeycomb micropillar lattices used in this work are fabricated from the planar microcavity using electron beam lithography followed by dry etching. Each lattice is composed of micropillars with equal radii $R_A = R_B = \SI{3.71}{\micro\meter}$, arranged with a center-to-center distance of $a=\SI{3.36}{\micro\meter}$. The resulting micropillar overlap leads to hybridization of the confined modes, and to the formation of energy bands. The band structure can be described by an effective tight-binding model, where the pillar size sets the on-site energy and their spatial overlap sets the hopping amplitude between neighboring sites.

\subsection{Tight-binding Hamiltonian in the ${\bm s}$-${\bm p}$ basis}

The $s$-$p$ honeycomb lattice considered in this paper can be described by the following tight-binding Hamiltonian:
\begin{align*}
    \hat{H} & = \hat{H}_s + \hat{H}_p + \hat{H}_{sp} \, ,\\
    \hat{H}_s & = 
    \sum_{i} \epsilon_{i,s} \, \hat{b}_{i,s}^\dagger \hat{b}_{i,s} 
    - \sum_{\langle i,j \rangle} t_{s,ij} \, \hat{b}_{i,s}^\dagger \hat{b}_{j,s} \, ,\\
    \hat{H}_p & = 
    \sum_{i}    \epsilon_{i,p_x} \, \hat{b}_{i,p_x}^\dagger \hat{b}_{i,p_x}
                + \epsilon_{i,p_y} \, \hat{b}_{i,p_y}^\dagger \hat{b}_{i,p_y} \\
    & - \sum_{\langle i,j \rangle} t_{p,ij} \, (\hat{\bm{b}}_{i,p}^\dagger \cdot \bm{e}_{ij})
    (\hat{\bm{b}}_{j,p} \cdot \bm{e}_{ij}) \, , \\
    \hat{H}_{sp} & = 
    \sum_{\langle i,j \rangle} t_{sp,ij} \, \hat{b}_{i,s}^\dagger (\hat{\bm{b}}_{j,p} \cdot \bm{e}_{ij}) \, .
\end{align*}
where $\hat{b}_{i,\sigma}$ ($\hat{b}_{i,\sigma}^\dagger$) denotes the annihilation (creation) operator in the $\sigma$ orbital at site $i$, $\epsilon_{i,\sigma}$ is the energy of the $\sigma$-orbital at site $i$, $t_{\sigma,ij}$ is the coupling amplitude between the orbitals $\sigma$ in site $i$ and in site $j$, $t_{sp,ij}$ is the coupling amplitude between orbital $s$ in site $i$ and orbital $p$ in site $j$, $\bm{e}_{ij}$ is the normalized vector in the direction of the $i$-$j$ link, and $\hat{\bm{b}}_{i,p} = \hat{b}_{i,p_x} \bm{e}_x + \hat{b}_{i,p_y} \bm{e}_y$. We note that this Hamiltonian neglects the transverse coupling between different $p$ orbitals, which has no significant impact on the topology of the system.

The lattice Bloch Hamiltonian is given by the following $6 \times 6$ matrix:
\begin{equation*}
    \hat{H}(\bm{k})=
    \begin{bmatrix}
        \hat{H}_s(\bm{k}) & \hat{H}_{sp}(\bm{k}) \\
        \hat{H}_{sp}(\bm{k})^\dagger & \hat{H}_{p}(\bm{k})
    \end{bmatrix}
    \, ,
\end{equation*}
with $\hat{H}_{s(p)} (\bm{k})$ the Bloch Hamiltonian of the $s(p)$ orbitals and  $\hat{H}_{sp} (\bm{k})$ is the $2 \times 4$ coupling block between them. In the basis $\{ \ket{s}, \ket{p_x}, \ket{p_y} \} \otimes \{ \ket{A}, \ket{B} \}$, these blocks take the explicit form:
\begin{align*}
    \hat{H}_s(\bm{k}) &= 
    \begin{bmatrix}
        \epsilon_s        &   -t_s \gamma_s(\bm{k}) \\
        - t_s \gamma_s(\bm{k})^* &  \epsilon_s
    \end{bmatrix}\, ,
    \\
    \hat{H}_p(\bm{k}) &= 
    \begin{bmatrix}
        \epsilon_p + \epsilon_{\text{el}}/2 & 0  & t_p\gamma^1_p(\bm{k})  &  t_p\gamma^2_p(\bm{k})       \\
        0   &   \epsilon_p - \epsilon_{\text{el}}/2 &   t_p\gamma^2_p(\bm{k}) & t_p\gamma^3_p(\bm{k})       \\
        t_p\gamma^1_p(\bm{k})^* &  t_p\gamma^2_p(\bm{k})^* & \epsilon_p + \epsilon_{\text{el}}/2 &   0                      \\
        t_p\gamma^2_p(\bm{k})^* & t_p\gamma^3_p(\bm{k})^*       & 0  &   \epsilon_p - \epsilon_{\text{el}}/2
    \end{bmatrix}\, ,
    \\
    \hat{H}_{sp}(\bm{k}) &= t_{sp}
    \begin{bmatrix}
        0 &   0   & -\gamma^1_{sp}(\bm{k})  &  -\gamma^2_{sp}(\bm{k}) \\
        \gamma^{1}_{sp}(\bm{k})^*  &  \gamma^{2}_{sp}(\bm{k})^*   & 0  &   0  
    \end{bmatrix} \, .
\end{align*}
Here, $\epsilon_{\text{el}}$ denotes the on-site energy splitting between the $p_x$ and $p_y$ orbitals within each site, which arises from a slight pillar ellipticity along the $y$-axis introduced during fabrication. The parameters $t_s, \, t_p, \,t_{sp}$ are the nearest-neighbor coupling amplitudes, and the off-diagonal coefficients are:
\begin{align*}
    \gamma_s(\bm{k}) &= \beta e^{-i\bm{k} \cdot \bm{\delta}_{1}}+e^{-i\bm{k} \cdot \bm{\delta}_{2}}+e^{-i\bm{k} \cdot \bm{\delta}_{3}}\, , 
    \\
    \gamma^1_p(\bm{k}) &= \frac{3}{4}(e^{-i \bm{k} \cdot \bm{\delta}_2}+e^{-i \bm{k} \cdot \bm{\delta}_3}) \, ,
    \\
    \gamma^2_p(\bm{k}) &= \frac{\sqrt{3}}{4}(e^{-i \bm{k} \cdot \bm{\delta}_2}-e^{-i \bm{k} \cdot \bm{\delta}_3}) \, ,
    \\
    \gamma^3_p(\bm{k}) &= \beta e^{-i \bm{k} \cdot \bm{\delta}_1} + \frac{1}{4}(e^{-i \bm{k} \cdot \bm{\delta}_2}+e^{-i \bm{k} \cdot \bm{\delta}_3}) \, ,
    \\ 
    \gamma^1_{sp}(\bm{k}) &= \frac{\sqrt{3}}{2}(e^{-i \bm{k} \cdot \bm{\delta}_2}-e^{-i \bm{k} \cdot \bm{\delta}_3}) \, ,
    \\
    \gamma^2_{sp}(\bm{k}) &= -\beta e^{-i \bm{k} \cdot \bm{\delta}_1} + \frac{1}{2}(e^{-i \bm{k} \cdot \bm{\delta}_2}+e^{-i \bm{k} \cdot \bm{\delta}_3}) \, ,
\end{align*}
where ${\bm \delta} _ i$ are the nearest-neighbor vectors defined in \cref{fig:Graphene}a. The parameter $\beta$ accounts for the uniaxial strain along the $y$-direction induced by a small displacement of the pillar centers introduced during fabrication. The parameters used in tight-binding simulations, given in units of $t_p$, are $\epsilon_s = 0,\: \epsilon_{p} = 5,\: \epsilon_{\text{el}} = 0.14,\: t_s = 0.2,\: t_p = 1,\: t_{sp} = 0.2,\: \beta = 1.06$. In our lattices, the $s$-$p$ energy splitting is typically $\epsilon_p - \epsilon_{s} \simeq \SI{2.5}{\milli\electronvolt}$.

\subsection{Experimental set-up}

A continuous-wave laser tuned to \SI{1.589}{\electronvolt}, approximately \SI{100}{\milli \electronvolt} above the energy of the honeycomb lattice bands, is focused onto the sample with a typical spot size of about \SI{30}{\micro\meter}. The sample positioning is optimized using a white-light illumination system that enables real-space visualization of the sample surface and precise alignment of the excitation spot onto the chosen lattice area.

The photoluminescence signal is collected through a high–numerical-aperture (${\rm N.A.}~=0.55$) aspheric lens of focal length $f_c = \SI{4.56}{\milli\meter}$, placed inside the cryostat. The resulting real-space image of the lattice emission is then focused onto a Hamamatsu SLM by a three-lens imaging system composed of converging lenses with focal lengths $f_1 = \SI{200}{\milli\meter}$, $f_2 = \SI{200}{\milli\meter}$, and $f_3 = \SI{400}{\milli\meter}$. The incident light is linearly polarized along the diagonal with respect to the SLM active axis using a half-wave plate and a polarizer. The light beam reflects on the SLM under a small incidence angle $\theta < 5^\circ$, and passes through an analyzer oriented cross-diagonally relative to the input polarizer. This polarizer–SLM–analyzer configuration allows spatially resolved control of both the amplitude and the phase of the reflected emission pattern.

The modulated signal is then optically Fourier-transformed and imaged onto the entrance slit of a spectrometer using three additional lenses with focal lengths $f_4 = \SI{500}{\milli\meter}$, $f_5 = \SI{100}{\milli\meter}$, and $f_6 = \SI{200}{\milli\meter}$. The last lens is mounted on a translation stage, enabling selection of any $k_y$ slice of the Fourier plane. The spectrally resolved Fourier-space images are recorded with an ANDOR Solis $1024\times1024$ CCD camera coupled to the spectrometer, providing a pixel size of $\SI{13.3}{\micro\meter} \times \SI{13.3}{\micro\meter}$ and an energy resolution of $\delta E = \SI{28}{\micro\electronvolt\per\pixel}$.

The SLM operation is controlled by grayscale input images, with gray levels ranging from 0 to 255, each corresponding to a specific local phase shift. The phase calibration is performed by sending a reference laser beam through the system and measuring the transmitted intensity as a function of the grayscale level of a uniform SLM image. Representative examples of SLM masks used in this work are shown in~\cref{fig:Masks}a–d. Overall, this setup enables measuring the spectrally resolved Fourier-space emission intensity for a series of phase masks applied \textit{via} the SLM.

\subsection{Fourier space intensity distribution after filtering using the SLM}

In this work, we develop a method to extract the Bloch eigenvectors of the $6 \times 6$ Hamiltonian in \cref{eq:6Hamiltonian}. The approach relies on directly controlling the amplitude and phase of the light emitted from different sub-regions of each lattice site, thereby modulating the Fourier-space intensity distribution according to \cref{eq:IntensitySpectra2}. The approximations leading to \cref{eq:IntensitySpectra2} are described below.

The honeycomb lattice studied in this work is a photonic structure etched from a planar semiconductor microcavity, which optical modes are governed by Maxwell equations. By restricting the analysis to the transverse plane, the transverse photonic modes can be described by a 2D Schr\"odinger equation for the electromagnetic field amplitude $\Psi^{2D}_{n,\bm{k}} (\bm{r})$:
\begin{equation*}
    \left[-\frac{\hbar^2}{2 m_c} \Delta + V(\bm{r}) \right] \Psi^{2D}_{n,\bm{k}} (\bm{r}) =  E_{n,\bm{k}} \Psi^{2D}_{n,\bm{k}} (\bm{r}) \, ,
\end{equation*}
where $n$ is the band index, $k$ the in-plane wave vector and $m_c = \SI{3.0e-35}{\kilo\gram}$ the photon effective mass extracted by fitting the parabolic dispersion of the unpatterned planar cavity. The potential $V(r)$ is a step-like honeycomb potential of amplitude $V_0 = \SI{600}{\milli\electronvolt}$, matching the geometry of the etched lattice.

Using an approach akin to the linear combination of atomic orbitals (LCAO), and restricting ourselves to the six lowest bands, the lattice eigenmodes can be expressed as linear combinations of six fragment orbitals $\phi^\sigma$ with $\sigma \in \{\left( A,1\right), \left( A,2\right),\left( A,3\right), \left( B,1\right), \left( B,2\right), \left( B,3\right)\}$:
\begin{equation*}
\Psi^{2D}_{n,\bm{k}} (\bm{r}) = 
\frac{1}{\sqrt{N_{\rm{tot}}}} \,  \sum_{j,\sigma} v^{\sigma}_{n,\bm{k}} \, e^{i \bm{k} \cdot \bm{r}^\sigma_j} \,  \phi^{\sigma}(\bm{r} - \bm{r}_{j}^\sigma) \, ,
\end{equation*}
where $N_{\rm{tot}}$ is the total number of unit cells, $\bm{r}_{j}^{\sigma}$ denotes the position of pillar center associated with $\phi^\sigma$ in unit cell $j$, and $v^{\sigma}_{n,\bm{k}}$ are the Bloch eigenvectors components in the $\{ \ket{sp^2} \}$ basis.

\begin{widetext}
Our approach for measuring the eigenstates consists of using the SLM to select the emission of small sectors of radius $R_m$ centered at the positions described in \cref{fig:PillarOrbitals}. In each sector, we apply a local amplitude attenuation or phase shift described by the complex coefficient $m^{ \sigma}$. After the SLM the filtered modes are given by:
\begin{align}
    \Psi^{2D}_{n,\bm{k} | m} (\bm{r}) &  = 
    \frac{1}{\sqrt{N_{\rm{tot}}}} \, 
    \sum_{j,\, \sigma} v^{\sigma}_{n,\bm{k}}\,
    m^{\sigma}\,
     e^{i \bm{k} \cdot \bm{r}^\sigma_j} \,
    \phi^{\sigma}(\bm{r} - \bm{r}_{j}^\sigma) \,
    \Theta \left( R_m - |\bm{r} - \bm{r}_j^{\sigma}-\bm{d}^{\sigma} | \right) \, ,
    \label{eq:2DWavefunctionRealSpace}
\end{align}
where $\Theta(R)$ denotes the Heaviside step function, equal to 1 for positive arguments and 0 otherwise. The vector $\bm{d}^\sigma$ points from the center of the lattice site to the center of the $\sigma$ sector selected by the SLM with length $|d^\sigma| = ({2-\sqrt{3}})a$, as shown in \cref{fig:PillarOrbitals}a. Importantly, in~\cref{eq:2DWavefunctionRealSpace}, we have assumed that $R_m$ has been chosen small enough compared to the pillar radius, such that within each sector,  one collects signal coming from one fragment orbital only, while the contributions from all neighboring orbitals vanish. In practice, we use in the experiments $R_m/R_A = 0.3$.

From this equation, we obtain the $k$-space wavefunction by Fourier transforming:
\begin{align*}
    \varphi_{n,{\bm k}|m}({\bm q}) & 
     = \frac{1}{\sqrt{N_{\rm tot}}} \sum_{j, \, \sigma} 
    m^\sigma \, v_{n,{\bm k}}^\sigma \, e^{i \bm{k} \cdot \bm{r}^\sigma_j}
    \iint_{\mathbb{R}^2} e^{ - i {\bm q} \cdot {\bm r}} \, \phi^{\sigma}(\bm{r} - \bm{r}_{j}^\sigma)
    \Theta \left( R_m - |\bm{r} - \bm{r}_j^\sigma-\bm{d}^\sigma | \right) \, d{\bm r}  \\
    & = \sum_{\sigma} m^\sigma \, v_{n,{\bm k}}^\sigma 
    \left( \frac{1}{\sqrt{N_{\rm tot}}}  \sum_j e^{i (\bm{k} - \bm{q}) \cdot \bm{r}^\sigma_j} \right)\,
    e^{-i\bm{q} \cdot \bm{d}^\sigma} \iint_{|\bm{r}|<R_m} e^{ - i {\bm q} \cdot {\bm r}} \, \phi^{\sigma}(\bm{r} + \bm{d}^\sigma) \, d{\bm r} \\
    & = \sqrt{N_{\rm tot}} \sum_{\sigma} m^\sigma \, v_{n,{\bm k}}^\sigma 
    S(\bm{k}-\bm{q}) \,
    e^{-i\bm{q} \cdot \bm{d}^\sigma} \tilde{\phi}^\sigma\vert_{R_m} (\bm{q}) \, ,
\end{align*}
where we have performed the change of variables $\bm{r} - \bm{r}_j^\sigma-\bm{d}^\sigma \rightarrow \bm{r}$, $S(\bm{k}-\bm q) = \frac{1}{N_{\rm tot}}  \sum_j e^{i (\bm{k} - \bm{q}) \cdot \bm{r}^\sigma_j}$ is a structure factor, and $\tilde{\phi}^\sigma\vert_{R_m}(\bm{q})$ is the Fourier transform of $\phi^{\sigma}({\bm r})$ truncated to $|{\bm r}-\bm{d}_\sigma| < R_m$.

\end{widetext}

\noindent For sufficiently large $N_{\rm tot}$, $S({\bm k} - {\bm q})$ is sharply peaked around ${\bm q} = {\bm k}$, leading to a highly directional emission along the ${\bm k}$ direction of amplitude:
\begin{equation*}
    \varphi_{n,{\bm k}|m}({\bm k})
     = \sqrt{N_{\rm tot}} \sum_{\sigma} m^\sigma \,  v^{\sigma}_{n,\bm{k}}  \, e^{-i\bm{q} \cdot \bm{d}^\sigma} \,\tilde{\phi}^{\sigma}\vert_{R_m}(\bm{k}) \, .
\end{equation*}
Because of the symmetries of the honeycomb lattice and of the $\ket{sp^2}$ orbitals, one finds that all $\tilde{\phi}^{\sigma}\vert_{R_m}(\bm{k}) \equiv \tilde{\phi}_{0}(\bm{k})$ are equal for any value of $\sigma$, so that we can finally write:
\begin{equation*}
    \varphi_{n,{\bm k}|m}({\bm k})
     = \sqrt{N_{\rm tot}} \tilde{\phi}_{0}(\bm{k})  \sum_{\sigma} m^\sigma \,  v^{\sigma}_{n,\bm{k}} \,  e^{-i {\bm k} \cdot \bm{d}_\sigma}\, ,
\end{equation*}
thus proving \cref{eq:IntensitySpectra2} of the main text.

\subsection{Measuring the Bloch eigenvectors}

Our method for measuring the lattice Bloch eigenstates relies on performing thirty-six intensity measurements $I_{\bm m}(E,{\bm k})$, under well-chosen phase configurations of the SLM encoded by the vector $\bm{m}$. In what follows, we show how the choice of configurations reported in the main text indeed allow us to measure all matrix elements of the form $\sum_{n=1}^6 \eta^n_{\bm{k}}(E) v^{\sigma}_{n,\bm{k}} {v^{\sigma'}_{n,\bm{k}}}^* $, and thus to reconstruct the density operator $\hat{\rho} (E,{\bm k})$. As mentioned in the main text, we define the vectors $\bm{m}^\sigma$ as unit vectors whose only nonzero component is the $\sigma$ component. We then perform i.~six measurements for all possible vectors $\bm{m}^\sigma$, ii.~fifteen measurements for all possible vectors $\bm{m}^\sigma + \bm{m}^{\sigma'}$ with $\sigma < \sigma'$, and iii.~fifteen measurements for all possible vectors $(\bm{m}^\sigma + i\bm{m}^{\sigma'})/\sqrt{2}$ with $\sigma < \sigma'$. The corresponding intensity distributions are given by:
\begin{widetext}
\begin{align}
     I_{{\bm m}^\sigma} (E,{\bm k})
     &= \sum_{n=1}^6 \eta^n_{\bm{k}}(E)
     \left | v^{\sigma}_{n,\bm{k}} \right | ^ 2 \, , \label{eq:Im} \\
     I_{{\bm m}^\sigma+{\bm m}^{\sigma'}} (E,{\bm k})
     &= \sum_{n=1}^6 \eta^n_{\bm{k}}(E) \left(
     \left | v^{\sigma}_{n,\bm{k}} \right | ^ 2 + \left | v^{\sigma'}_{n,\bm{k}} \right | ^ 2 + 2 \Re \left[ e^{-i {\bm k} \cdot ({\bm d}^\sigma - {\bm d}^{\sigma '} )} v^{\sigma}_{n,\bm{k}} {v^{\sigma'}_{n,\bm{k}}}^* \right] \right)\, , \label{eq:Im+m'} \\
     I_{({\bm m}^\sigma+i{\bm m}^{\sigma'})/\sqrt{2}} (E,{\bm k})
     &= \frac{1}{2}\sum_{n=1}^6 \eta^n_{\bm{k}}(E) \left(
     \left | v^{\sigma}_{n,\bm{k}} \right | ^ 2 + \left | v^{\sigma'}_{n,\bm{k}} \right | ^ 2 + 2 \Im \left[ e^{-i {\bm k} \cdot ({\bm d}^\sigma - {\bm d}^{\sigma '} )} v^{\sigma}_{n,\bm{k}} {v^{\sigma'}_{n,\bm{k}}}^* \right] \right) \, . \label{eq:Im+im'}
\end{align}
From \cref{eq:Im}, one directly obtains the diagonal matrix elements of $\hat{\rho} (E,{\bm k})$. Furthermore, the following linear combination of \cref{eq:Im}, \cref{eq:Im+m'} and \cref{eq:Im+im'} allows us to determine the off-diagonal elements of $\hat{\rho} (E,{\bm k})$:
\begin{equation*}
    \frac{1}{2} e^{i {\bm k} \cdot ({\bm d}^\sigma - {\bm d}^{\sigma'} )} \left[ I_{{\bm m}^\sigma + {\bm m}^{\sigma'}} - I_{{\bm m}^\sigma} - I_{{\bm m}^{\sigma'}} + i \left( 2 I_{({\bm m}^\sigma + i {\bm m}^{\sigma'})/\sqrt{2}} - I_{{\bm m}^\sigma} - I_{{\bm m}^{\sigma'}} \right) \right] =  \sum_{n=1}^6 \eta^n_{\bm{k}}(E) v^{\sigma}_{n,\bm{k}} {v^{\sigma'}_{n,\bm{k}}}^* \, .
\end{equation*}
This procedure thus provides the full reconstruction of $\hat{\rho} (E,{\bm k})$.
\end{widetext}

From the expression of $\hat{\rho}(E,\bm{k})$, it is clear that its eigenvectors directly yield the Bloch eigenstates $\ket{v_{n,\bm{k}}}$. For a given $\bm{k}$, we obtain multiple density matrices, as many as the pixel number along the energy axis of the spectrometer. We select only those with a sufficiently large signal-to-noise ratio, and then perform a joint diagonalization using the JADE algorithm, giving us the Bloch eigenvectors. The density matrix method used to recover the Bloch eigenvectors can also be used to track the energy of each band with high precision. Indeed, once the eigenvectors are known, it is possible to separate the Lorentzian profile $\eta^n_{\bm{k}}(E)$ in each band. To do so, we compute ${\rm Tr} \left( \ket{u_{n,\bm{k}}} \! \bra{u_{n,\bm{k}}} \hat{\rho} (E, \bm{k}) \right)$, and then fit the resulting profile with a single Lorentzian curve. The result of such a procedure can be found in \cref{fig:ResolvedSpectra}, where, for a given value of ${\bm k}$, the total measured intensity ${\rm Tr} \left( \hat{\rho} (E) \right)$ as well as the projected components ${\rm Tr} \left( \ket{u_{n,\bm{k}}} \! \bra{u_{n,\bm{k}}} \hat{\rho} (E, \bm{k}) \right)$ are displayed. The different Lorentzian lines are well resolved, enabling us to accurately determine the energy of each band with sub-linewidth precision.

\subsection{Numerical computation of the non-Abelian QGT and Euler class}

The Euler class is defined in systems with $\mathcal{C}_2 \mathcal{T}$ or $\mathcal{P} \mathcal{T}$ symmetry, which guarantees that the Bloch Hamiltonian can be rendered real using the Takagi factorization. As a first step toward computing the non-Abelian QGT and the Euler class, we perform a basis transformation to express the measured Bloch Hamiltonian in a real form. This is achieved using the transformation matrix:
\begin{equation*}
    V = \sqrt{
        \begin{bmatrix}
            0  & 1  & 0  & 0  &  0 & 0  \\
            1  & 0  & 0  & 0  &  0 & 0  \\
            0  & 0  & 0  & 0  & -1 & 0  \\
            0  & 0  & 0  & 0  & 0  & -1 \\
            0  & 0  & -1 & 0  & 0  & 0  \\
            0  & 0  & 0  & -1 & 0  & 0
        \end{bmatrix}}
        \, .
\end{equation*}
After the transformation, a residual imaginary component may persist due to experimental uncertainties. This component is discarded, yielding a purely real Hamiltonian. We then apply a smoothing to this Hamiltonian using a Gaussian kernel of width $\sigma = 2 \: \text{pixels}$.

\medskip

\paragraph*{\textbf{\textit{Computing the Euler curvature.}}}
We diagonalize the real-valued Hamiltonian, which yields the corresponding real eigenvectors $\ket{r_{n,{\bm k}}}$. We numerically fix the gauge of these eigenvectors so that it is smooth everywhere except along straight lines connecting the band nodes~\cite{Ahn2019,Bouhon2019}. We pair each node with its closest neighbor, starting with the shortest overall distance. These lines, where the sign of the eigenvectors flips, are the DSs connecting each pair of Dirac cones. To compute the Euler curvature (the non-Abelian Berry curvature), we use the fact that the Euler curvature for a given pair of eigenvectors $\text{Eu}\left(\ket{r_{n,\bm{k}}}, \, \ket{r_{n+1,\bm{k}}}\right)$ is equal to the (Abelian) Berry curvature of a complexified band with wavefunction $\ket{s_{n,n+1,{\bm k}}} =  \left( \ket{r_{n,\bm{k}}} + i \ket{r_{n+1,\bm{k}}} \right) / \sqrt{2}$~\cite{Bouhon2019}. We evaluate the Berry curvature $B_{n,n+1}(\bm{k})$ of this complexified band by constructing a plaquette around ${\bm k}$ with states $\ket{s1} = \ket{s_{n,n+1,\bm{k}}}$, $\ket{s2} = \ket{s_{n,n+1,\bm{k}+dk \mathbf{e_x}}}$, $\ket{s3} = \ket{s_{n,n+1,\bm{k}+dk \mathbf{e_x}+dk \mathbf{e_y}}}$, $\ket{s4} = \ket{s_{n,n+1,\bm{k}+dk \mathbf{e_y}}}$. The curvature is finally obtained from the so-called four-point formula:
\begin{equation*}
    B_{n,n+1}(\bm{k})   = \arg \left[\braket{s1|s2}\braket{s2|s3}\braket{s3|s4}\braket{s4|s1}\right] \, .
\end{equation*}

\medskip

\paragraph*{\textbf{\textit{Computing the Euler class.}}} We first define a patch within the Brillouin zone that does not contain any adjacent nodes, ensuring that the eigenstates of the target bands are well defined throughout this region. We then fix the gauge so that it remains smooth everywhere except along a straight line connecting the principal nodes within the patch: any Dirac string in between adjacent band pairs can indeed be pushed outside of the patch~\cite{Breach2024}. We compute the Euler curvature over the entire patch following the procedure described above, and evaluate the Euler connection along the patch boundary using a standard numerical derivative scheme. The patch Euler invariant is then obtained using \cref{eq:Eulerpatch} of the main text. In all cases, we find well-quantized values with deviations below $10\%$. The main source of uncertainty is numerical, arising from the precise choice of the patch boundaries and from the smoothing applied to the Hamiltonian.

\medskip

\paragraph*{\textbf{\textit{Computing the non-Abelian quantum metric.}}}
The non-Abelian QGT can be generalized from the single-band geometric tensor by considering $N$-band subsets. In this work we focus on two band subspaces indexed $\{\alpha,\beta\}$, although we emphasize that the calculation can be extended to larger subsets since we perform a full six-band eigenstate tomography. We first compute the real eigenvectors by fixing the gauge such that the DSs are oriented parallel to $k_y$. This choice ensures that the eigenvector components remain continuous everywhere except along $k_y$, thereby minimizing numerical errors in the evaluation of eigenstate derivatives. The components of the fourth-order QGT tensor are given by:
\begin{equation*}
G^{\alpha,\beta}_{i,j}=\langle \partial_i r_{\alpha,\bm{k}}|(1-\hat{P})| \partial_j r_{\beta,\bm{k}}\rangle \, ,
\end{equation*}
where $(i,j)$ indexes $k_x$ or $k_y$, and the operator $\hat{P} = |r_{\alpha,{\bm k}}\rangle \langle r_{\alpha,{\bm k}}| + |r_{\beta,{\bm k}}\rangle \langle r_{\beta, {\bm k}}|$ is the projector on the $(\alpha,\beta)$ subset of bands . The eigenstate derivatives are computed numerically using finite differences.

The non-Abelian quantum metric $g^{\alpha,\beta}_{i,j}$ is obtained by evaluating the symmetric part of this real QGT tensor: $g^{\alpha,\beta}_{i,j} = \frac{1}{2}\left(G^{\alpha,\beta}_{i,j}+G^{\alpha,\beta}_{j,i}\right)$. All components of the quantum metric are plotted in \cref{fig:QGT_1} for the $(3,4)$ and $(4,5)$ subsets of bands. We note that off-diagonal components $G^{\alpha,\beta}_{j,i}$ with $\alpha\ne\beta$ are gauge-dependent quantities, so that the discontinuities in the eigenstates (the DSs) explicitly appear in these components and may introduce numerical artifacts in their vicinity. Also, the QGT is inherently ill-defined at the location of adjacent nodes. This is indicated by gray points in the plots of ${\rm Tr} (g_{i,j}^{\alpha,\beta})$ in \cref{fig:Heff}e-h. Finally, we point out that the non-Abelian Berry curvature coincides with the anti-symmetric part of the QGT, $w^{\alpha,\beta}_{i,j}=\frac{1}{2}\left(G^{\alpha,\beta}_{i,j}-G^{\alpha,\beta}_{j,i}\right)$, which is proportional to the Euler curvature $\mathrm{Eu}_{\alpha,\beta}$ in $\mathcal{C}_2 \mathcal{T}$ symmetric systems, thus providing another way to compute the Euler curvature.

\subsection{Numerical computation of two-band effective Hamiltonians}

Following~\cite{Milicevic2019}, we compute effective two-band Hamiltonians from the measured six-band Hamiltonian. For a subset of two touching bands, we identify a point $\bm{k}_{0}$ in reciprocal space close to the studied degeneracies and express the Hamiltonian in the basis of the eigenstates at $\bm{k}_{0}$. As an example, for the pairs of bands $(3,4)$ or $(5,6)$, we select the $\Gamma$ point ($\bm{k}_{0} = \bm{0}$), as the reference point around which the tilted Dirac cones are centered. We first compute the eigenstates $\ket{u_{i,{\bm k}_0}}$ of $\hat{H} ({\bm k}_0)$ and reorder them to form a basis $\left \{ \ket{u_{i,{\bm k}_0}} \right \}_{i=1}^6$, where the first two vectors are the ones involved in the band touching point. In order to continuously select the gauge, we impose the sign of one of the component of the eigenstate. Namely, for $s$ ($p$) band eigenvectors, the $s_A$ ($p_{x, A}$) component is set to be real positive. This common gauge choice allows to compare phase windings of effective Hamiltonians computed at different points.

Written in this new basis, the total $6\times6$ Hamiltonian  $\tilde{H}(\bm{k})$ decomposes into four distinct blocks:
\begin{equation*}
    \tilde{H}(\bm{k}) = 
    \begin{bmatrix}
        \hat{H}_+(\bm{k}) &   \hat{C}(\bm{k}) \\
        \hat{C}^\dagger (\bm{k})&   \hat{H}_-(\bm{k})  
    \end{bmatrix}
\end{equation*}
with $\hat{H}_+(\bm{k})\in \mathcal{M}_{2\times2}$, $\hat{H}_-(\bm{k})\in \mathcal{M}_{4\times4}$ and $\hat{C}(\bm{k})\in \mathcal{M}_{2\times4}$. The effective two-band Hamiltonian is given at first order by:
\begin{equation*}
\hat{H}_{\rm eff}(\bm{k}) = \hat{H}_{+}(\bm{k})+\hat{C}(\bm{k})(E-\hat{H}_{-}(\bm{k}_{0}))^{-1}\hat{C}^\dagger(\bm{k}) \, ,
\end{equation*}
where $E$ is the energy of one of the two studied bands at $\bm{k}_0$ (note that although the two eigenvalues at ${\bm k}_0$ may not strictly coincide, this choice has almost no impact on the obtained $\hat{H}_{\rm eff}$). In \cref{eq:EffectiveHamiltonian} of the main text, the obtained $\hat{H}_{\rm  eff}$ is decomposed on the Pauli matrix basis and can be represented on the Bloch sphere. Its poles correspond to the eigenvectors of $\hat{H}(\bm{k}_{0})$ of the two bands of interest. Around the band nodes, the vector $\lambda(\bm{k})$ mostly belongs to an orbital plane. In this plane, $\lambda(\bm{k})$ is parametrized by an azimuthal angle $\varphi$ plotted in \cref{fig:Heff} and in \cref{fig:BlochSphere}.


\section*{Data availability}

All datasets generated and analysed during this study are available upon request from the corresponding author.

\section*{Code availability}

All codes generated during this study are available upon request from the corresponding author.

\begin{acknowledgments}

This work was supported by the European Research Council (ERC) under the European Union's Horizon 2020 research and innovation programme through the Starting Grant ARQADIA (grant agreement no. 949730) and under the Horizon Europe programme (project ANAPOLIS, grant agreement no. 101054448), by the RENATECH network and the General Council of Essonne, by the Paris Ile-de-France R\'egion via DIM SIRTEQ and DIM QUANTIP, by the Plan France 2030 through the project QUTISYM ANR23-PETQ-0002, by the ANR project Ngauge (ANR-24-CE92-0011). R.-J.S.~acknowledges funding from an EPSRC ERC underwrite grant EP/X025829/1, and a Royal Society exchange grant IES/R1/221060. F.N.\"U.~acknowledges support from the Royal Society Grant~URF/R1/241667, the Simons Investigator Award (Grant No.~511029) and Trinity College Cambridge. The work of R.-J.S.~and F.N.\"U.~was performed in part at Aspen Center for Physics, which is supported by National Science Foundation grant PHY-2210452.

\end{acknowledgments}

\section*{Author contributions} 

M.G. designed the lattices and developed the experimental method for orbital tomography. M.G. and C.B. conducted the experiments, performed the data analysis, wrote the codes used for data processing and performed all numerical simulations. S.R. designed the sample structure and coordinated the sample fabrication together with M.M. and A.L. for the molecular beam epitaxy growth, M.G. for the lithography mask design, L.L.G. for the electron-beam lithography, and A.H. and I.S. for etching the structures. F.N.\"U.~and R.-J.S.~introduced the connections to the multi-gap topology and provided the theoretical background. S.R., J.B., M.G., C.B., F.N.\"U., and R.-J.S. took part in scientific discussions throughout the project. M.G. and C.B. prepared the figures for the manuscript, S.R., J.B., M.G., C.B., F.N.\"U., and R.-J.S. wrote the manuscript. S.R. conceived and supervised the project.


\section*{Competing interests}

The authors declare no competing interests.

\clearpage

\setcounter{figure}{0}
\makeatletter
\renewcommand{\fnum@figure}{Extended Data Fig.~\thefigure}
\makeatother


\begin{figure*}[t]
    \centering
    \includegraphics[width=\textwidth]{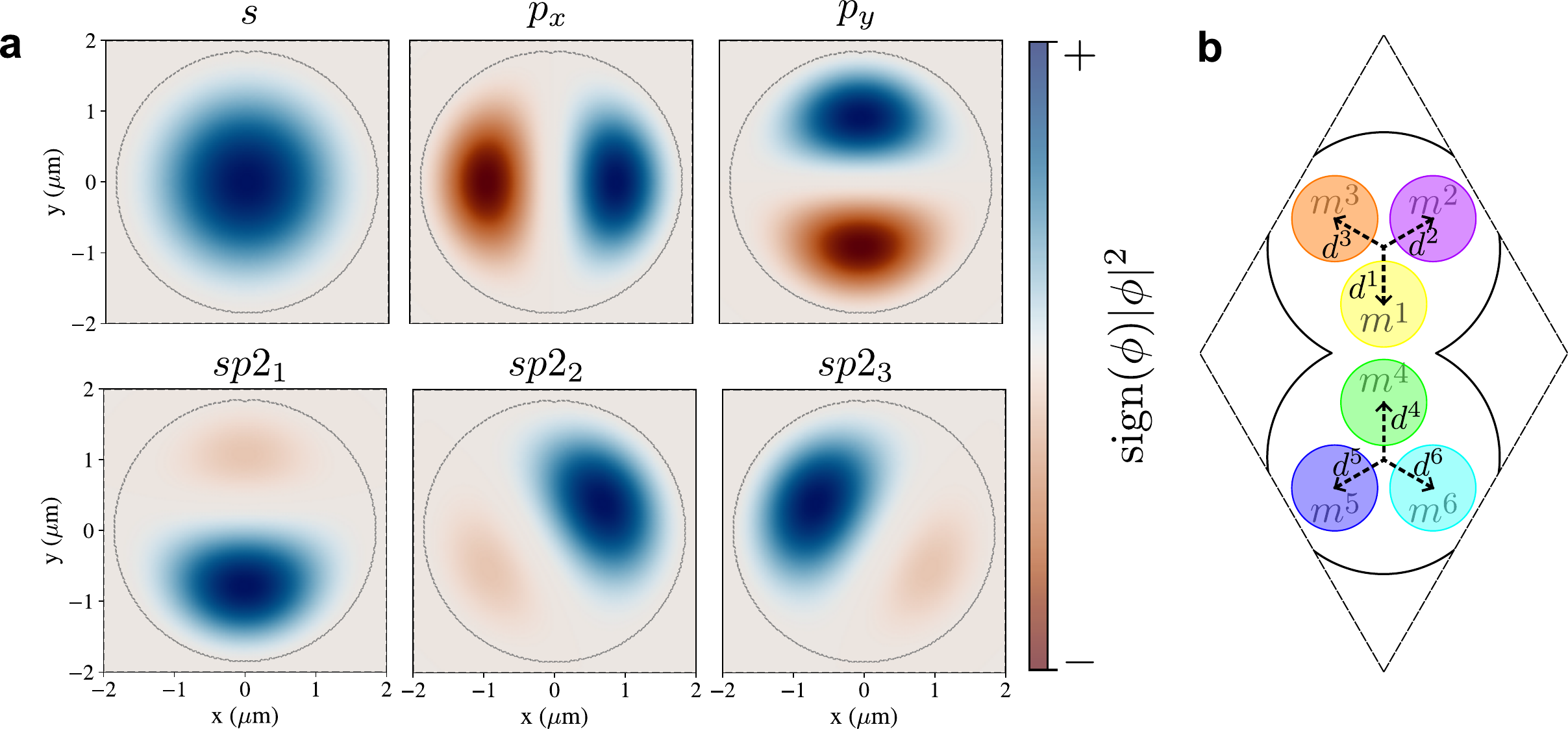}
    \caption{
    \textbf{a}.~Signed modulus square of the modes in a circular pillar in the $\{ \ket{s}, \ket{p_x}, \ket{p_y} \}$ basis (top panels), and in the $\{ \ket{sp^2_{\sigma}} \}$ basis (bottom panels). These modes are computed solving the 2D Schr\"odinger equation.
    \textbf{b}.~Representation of the unit cell sectors chosen to select the main lobe of the $\ket{sp^2_\sigma}$ orbitals using the SLM. Each sector is offset from the pillar center by a vector $d^\sigma$ of length $|d^\sigma| = ({2-\sqrt{3}})a$ aligned along the segment connecting the pillar to one of its neighbors.
    }
    \label[extfig]{fig:PillarOrbitals}
\end{figure*}

\begin{figure*}[t]
    \centering
    \includegraphics[width=\textwidth]{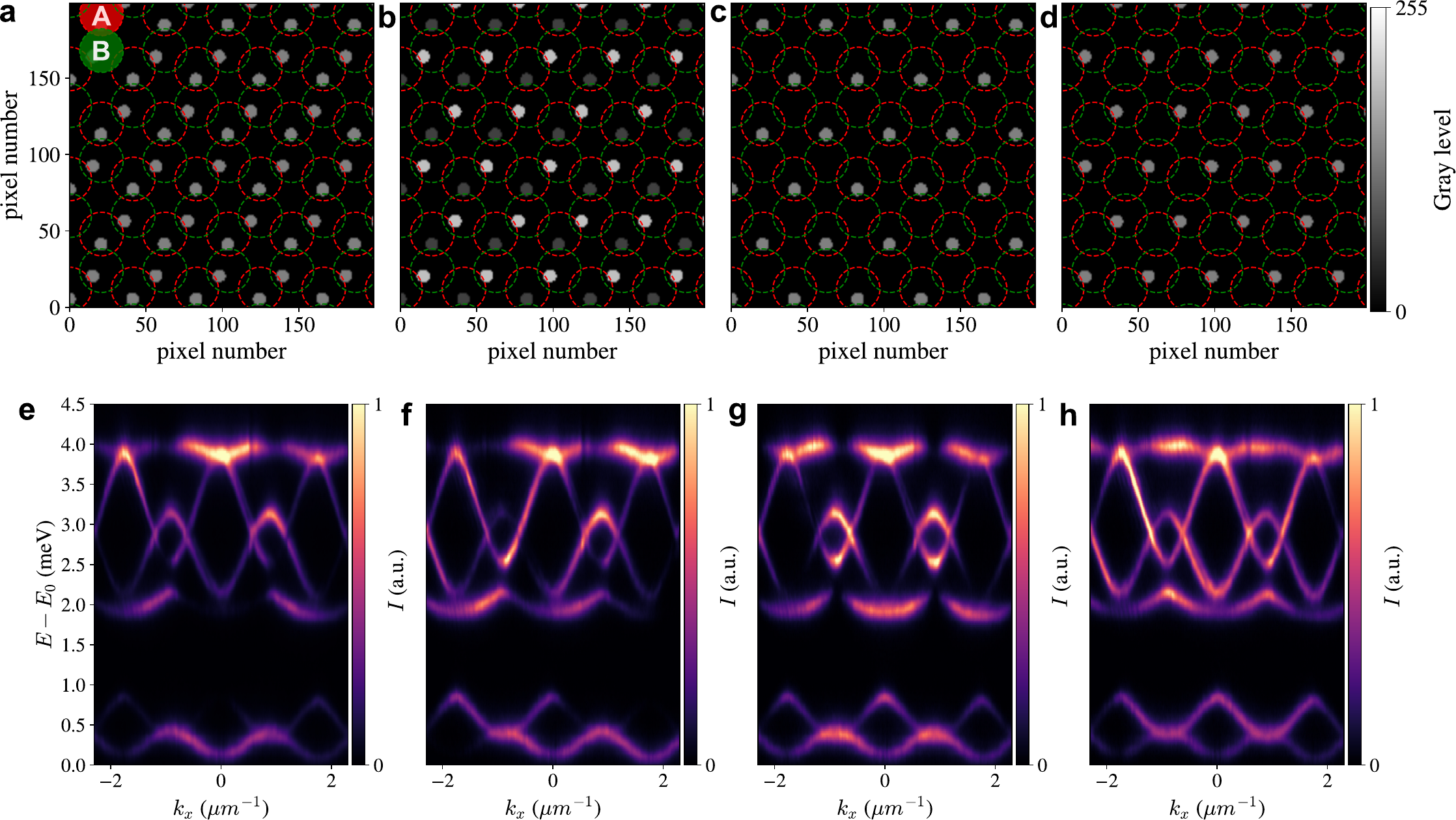}
    \caption{\textbf{a-d}:~Exemplary SLM input masks used to encode four different configurations of the ${\bm m}$ vector:
    \textbf{a}.~$\bm{m} = [1,0,0,0,1,0]$,
    \textbf{b}.~$\bm{m} = [1,0,0,0,i,0]/\sqrt{2}$,
    \textbf{c}.~$\bm{m} = [1,0,0,0,0,0]$ and
    \textbf{d}.~$\bm{m} = [0,0,0,0,1,0]$.
    \textbf{(e-h)}: Fourier space emission measured along $k_y = 0$ when applying the four masks shown in \textbf{a}-\textbf{d}. We clearly observe that the choice of $\bm{m}$ alters the Fourier space intensity distribution, due to the modification of the interference conditions between modes.}
    \label[extfig]{fig:Masks}
\end{figure*}

\begin{figure*}[t]
    \includegraphics[width=\textwidth]{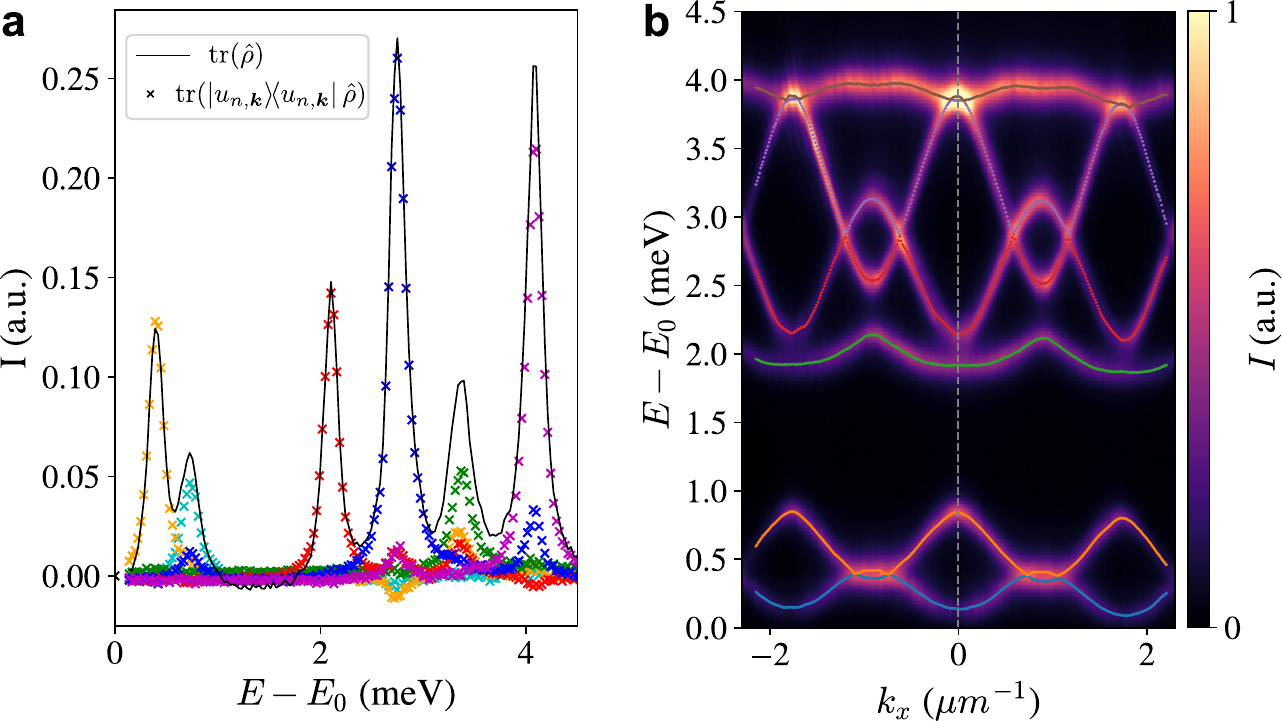}
    \caption{
    Determining the band energies with sub-linewidth precision.
    \textbf{a}. Representative energy spectrum at $\boldsymbol{k}=0$. The gray line shows the total intensity ${\rm Tr}[\hat{\rho}(E,\boldsymbol{k}=0)]$, while the colored symbols represent the projected spectra ${\rm Tr}[\ket{u_{n,\boldsymbol{k}}}\!\bra{u_{n,\boldsymbol{k}}}\hat{\rho}(E,\boldsymbol{k}=0)]$. Each projected spectrum is fitted with a Lorentzian profile (not shown) to determine the corresponding band energy with sub-linewidth accuracy.
    \textbf{b}.~Spectral intensity map ${\rm Tr}[\hat{\rho}(E,k_x,k_y=0)] = \sum_{\sigma=1}^6 I_{\boldsymbol{m}^\sigma}(E,k_x,k_y=0)$ showing the band dispersion along $k_x$ for fixed $k_y=0$. This representation eliminates interference effects and enhances visibility of all bands. The colored lines show the fitted peak energies determined as in panel (a). 
    }
    \label[extfig]{fig:ResolvedSpectra}
\end{figure*}

\begin{figure*}[t]
    \includegraphics[width=0.7\textwidth]{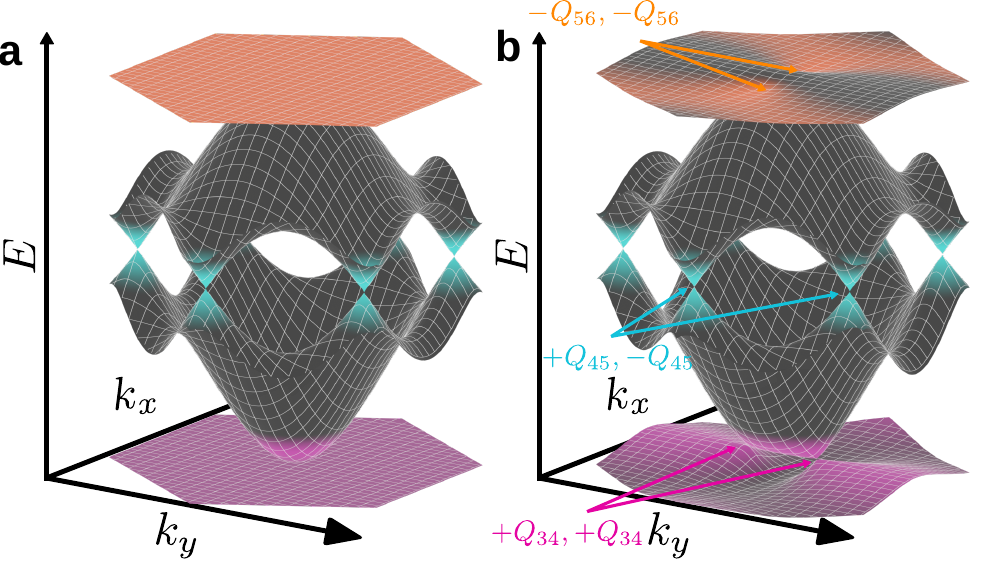}
    \caption{
    \textbf{a}.~Band dispersion computed solving the tight-binding Hamiltonian using the following parameters (in units of $t_p$): $\epsilon_s = 0,\: \epsilon_{p} = 5,\: \epsilon_{\text{el}} = 0$ (no ellipticity), $\: t_s = 0.2,\: t_p = 1,\: t_{sp} = 0.2,\: \beta = 1$ (no strain).
    \textbf{b}.~Same calculation as in \textbf{a}, including an ellipticity ($\epsilon_{\text{el}} = 0.14$) and a uniaxial strain along $y$ ($\beta = 1.06$). The arrows point towards the band nodes, each labeled by its generalized quaternion charge.
    }
    \label[extfig]{fig:3Ddispersions}
\end{figure*}

\begin{figure*}[t]
    \includegraphics[width=\textwidth]{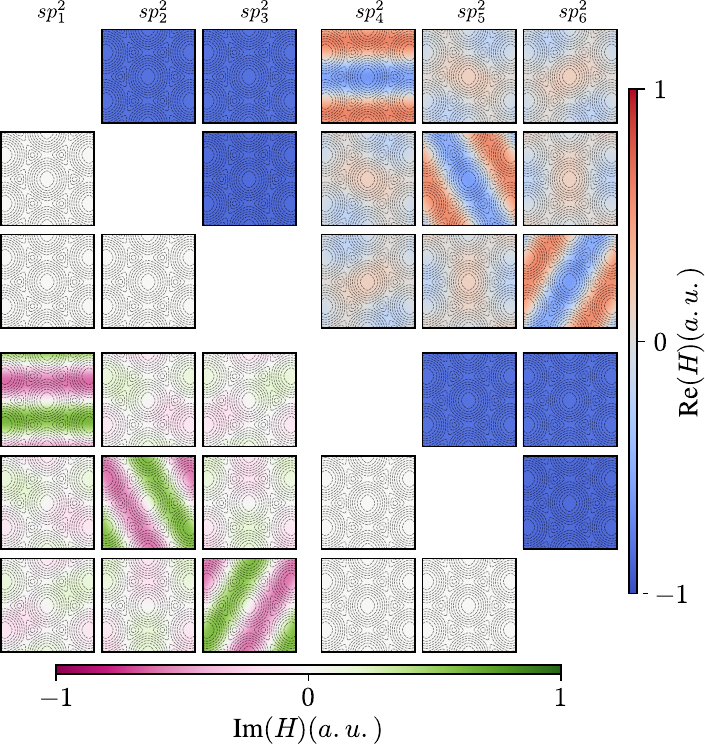}
    \caption{Off-diagonal matrix elements of the tight-binding Hamiltonian in \cref{eq:6Hamiltonian}, represented in the $\left\{ \ket{sp^2_{\sigma}} \right\}$ basis as a function of $k_x$ and $k_y$. Each panel shows the real (upper triangle) and imaginary (lower triangle) part of one of the Hamiltonian components. In each panel, dashed lines show iso-energy contours of the difference between bands 5 and 4. The parameters used (in units of $t_p$) are: $\epsilon_s = 0,\: \epsilon_{p} = 5,\: \epsilon_{\rm el} = 0.14,\: t_s = 0.2,\: t_p = 1,\: t_{sp} = 0.2,\: \beta = 1.06$. }
    \label[extfig]{fig:TBspHamiltonian}
\end{figure*}

\begin{figure*}[t]
    \includegraphics[width=\textwidth]{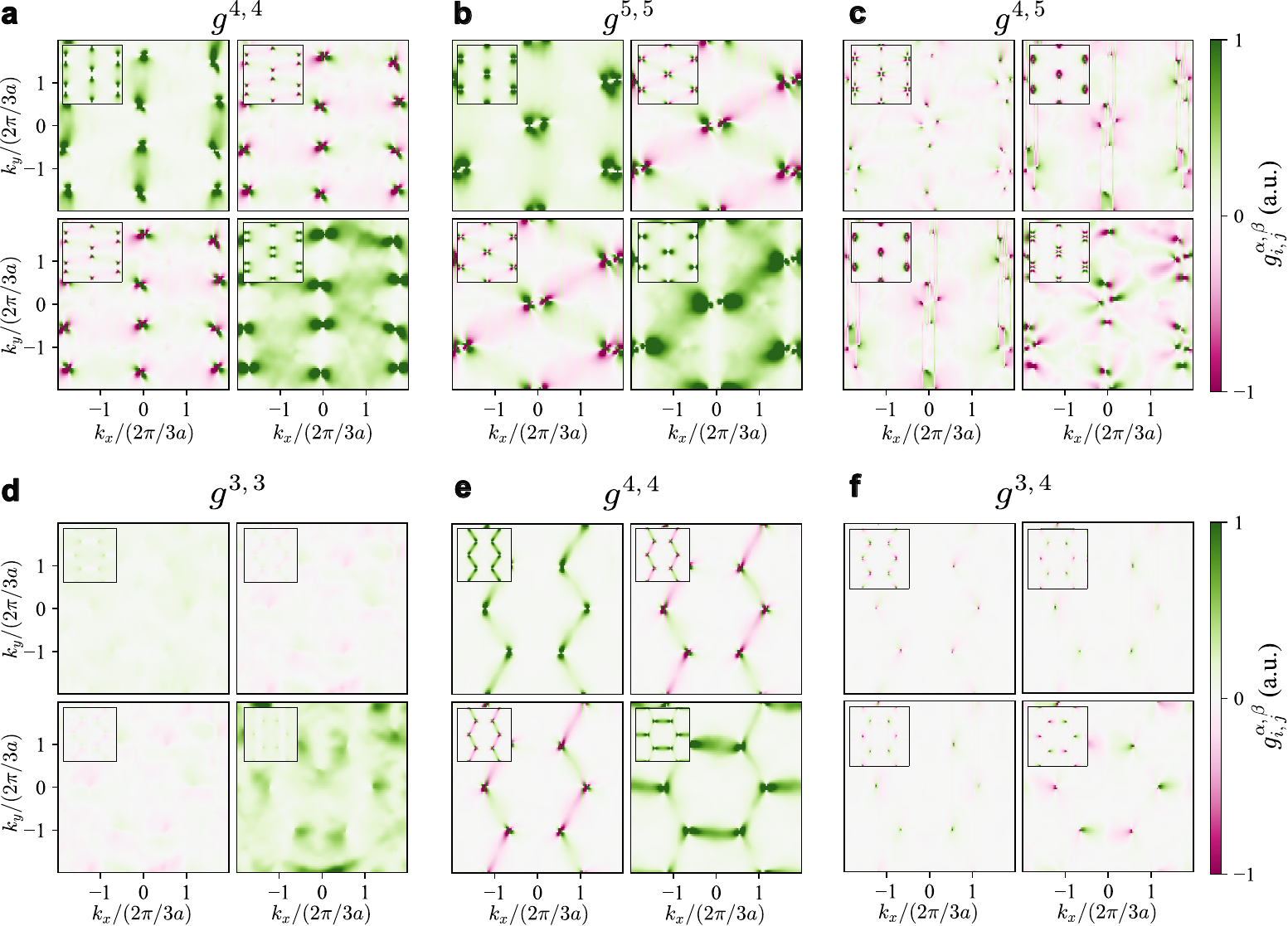}
    \caption{Components of the measured non-Abelian quantum metric $g^{\alpha,\beta}_{i,j}$ for \textbf{a}-\textbf{c}.~the subset of bands $(4, 5)$, and \textbf{d}-\textbf{f}.~the subset of bands $(3, 4)$. The $g^{\alpha, \beta}$ tensor components are plotted in the following order: $g^{\alpha, \beta}_{x, x}$ (top left), $g^{\alpha, \beta}_{x, y}$ (top right), $g^{\alpha, \beta}_{y, x}$ (bottom left), and $g^{\alpha, \beta}_{y, y}$ (bottom right). The insets show the result of tight-binding simulations with the following parameters (in units of $t_p$): $\epsilon_s = 0,\: \epsilon_{p} = 5,\: \epsilon_{\rm el} = 0.14,\: t_s = 0.2,\: t_p = 1,\: t_{sp} = 0.2,\: \beta = 1.06$.}
    \label[extfig]{fig:QGT_1}
\end{figure*}

\begin{figure*}[t]
    \includegraphics[width=\textwidth]{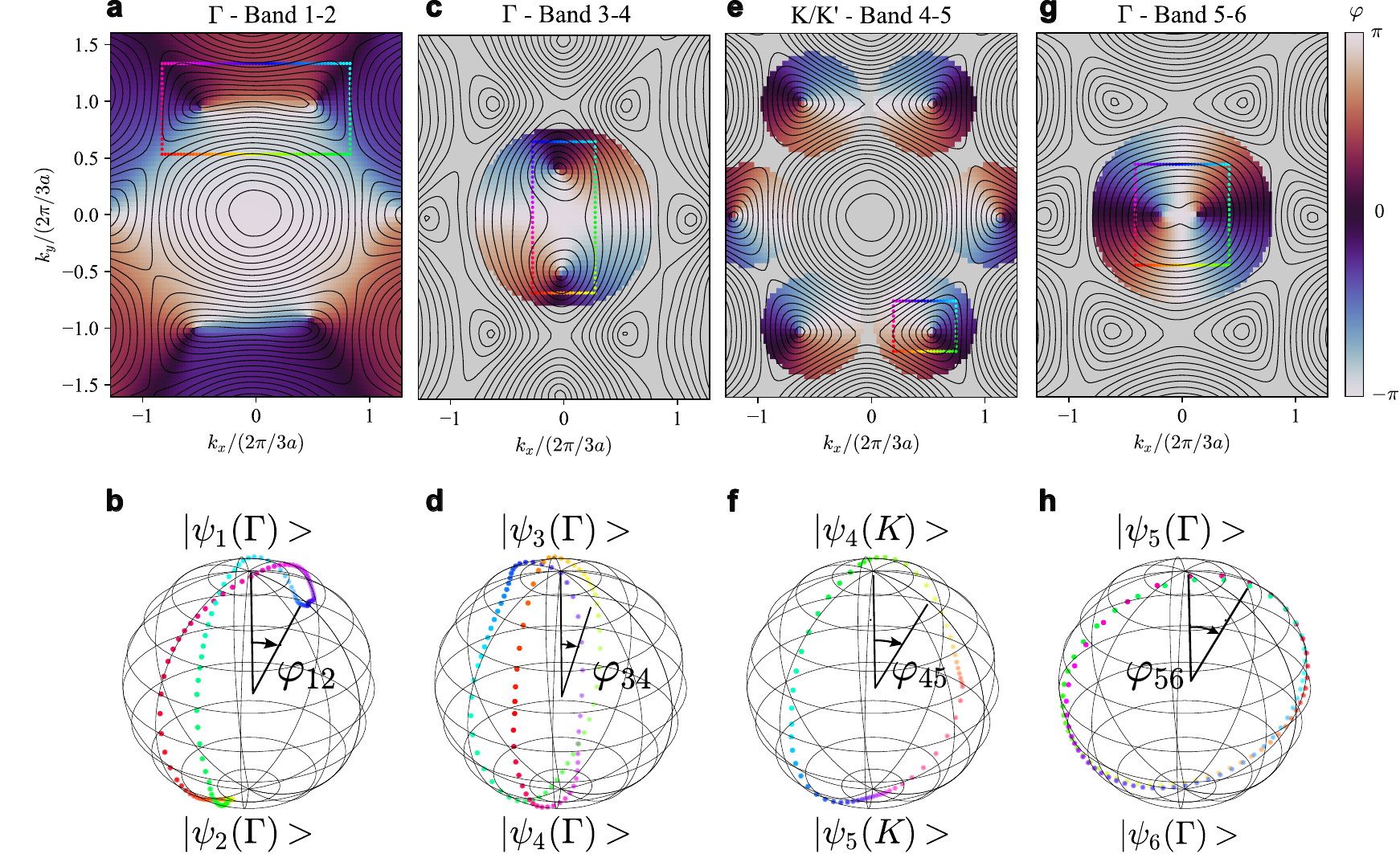}
    \caption{\textbf{a,b,c,d}.~Effective two-band Hamiltonians around selected ${\bm k_0}$ points (indicated above each panel). The color map represents the azimuthal angle $\varphi_{n,n+1}$ of the vector $\bm{\lambda}(\bm{q})$ in the orbital plane defined in the text, plotted over the $\boldsymbol{k}$-space regions where the two-band approximation holds. In all panels dashed lines show iso-energy contours of the energy difference between the two considered bands.
    \textbf{e,f,g,h}.~Trajectories of the vector $\bm{\lambda}(\bm{q})$ on the Bloch sphere corresponding to (a,b,c,d) when $\bm{q}$ varies along the contours encircling band nodes represented on the panel above each sphere. Color coding on the contours and corresponding trajectories are the same.}
    \label[extfig]{fig:BlochSphere}
\end{figure*}

\begin{figure*}[t]
    \includegraphics[width=0.7\textwidth]{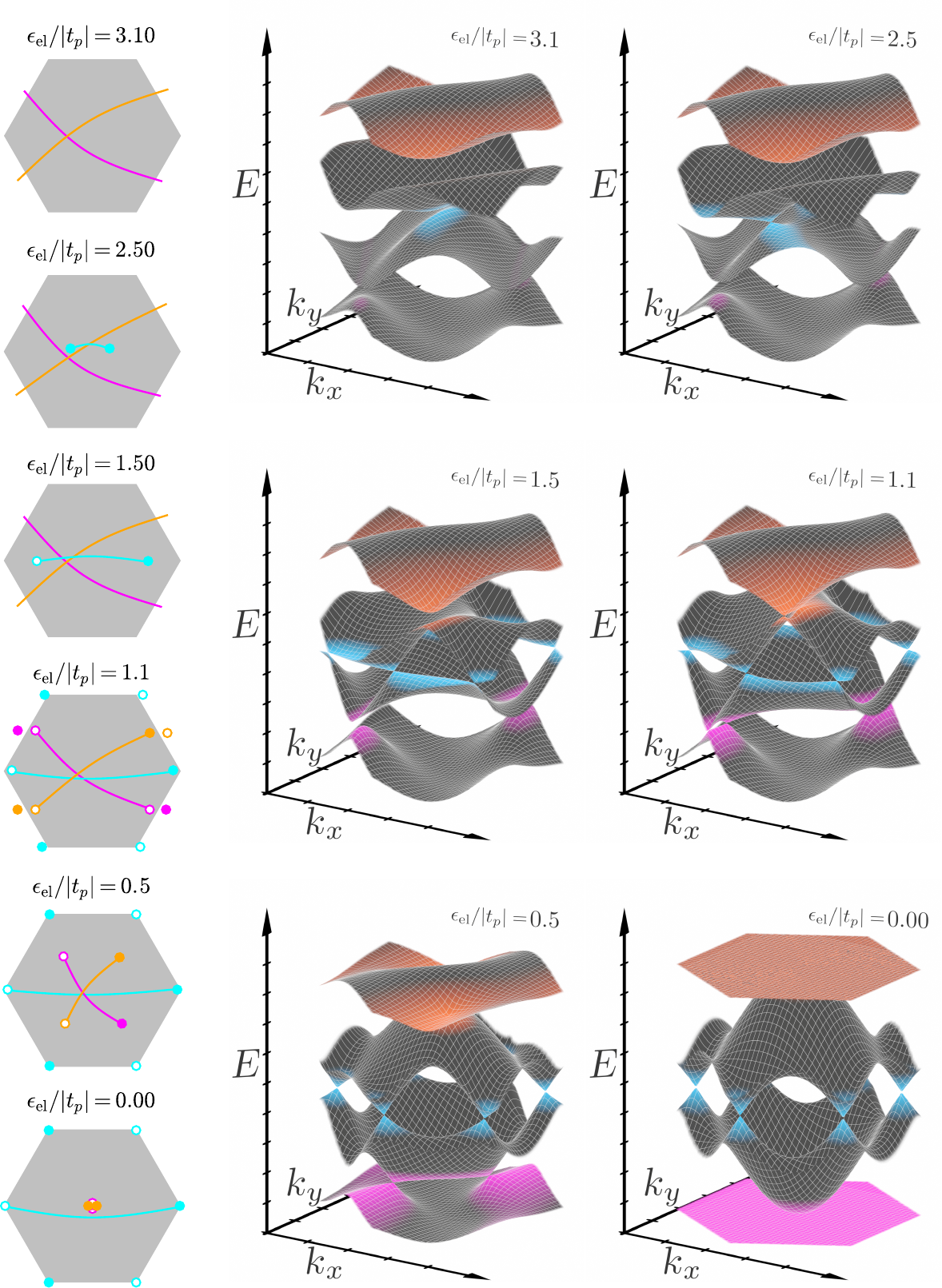}
    \centering
  \caption{Snapshot of a braiding protocol continuously tuning the ellipticity of the sites between $\epsilon_{el}/ |t_p|=3.10$ and $\epsilon_{el}/ |t_p|=0.00$, and a rotation of the ellipse axis 45$^\circ$ from the x-axis. A film of this braiding protocol can be found in the supplementary information.
    \textbf{(left column)}:~Overall topological configuration in the first Brillouin zone for six values of $\epsilon_{el}/ |t_p|$ indicated in the figure. We show the node positions with positive (full circle) or negative (empty circles) generalized quaternion charge and the DSs locations (solid lines). The colors encode the different gaps as in \cref{fig:Graphene}c.
    \textbf{(center and right columns)}:~Band dispersions computed using the tight binding Hamiltonian for different values of $\epsilon_{el}/ |t_p|$ indicated in the figure and a rotation of the ellipse axis 45$^\circ$ from the x-axis. The total band amplitudes have been normalized for clarity. Other tight binding parameters are (in units of $t_p$): $\epsilon_s = 0,\: \epsilon_{p} = 5,\: t_s = 0.2,\: t_p = 1,\: t_{sp} = 0.2,\: \beta = 1$.}
    \label[extfig]{fig:ExtdBraiding}
\end{figure*}

\newpage
\clearpage


\begin{thebibliography}{60}%
\makeatletter
\providecommand \@ifxundefined [1]{%
 \@ifx{#1\undefined}
}%
\providecommand \@ifnum [1]{%
 \ifnum #1\expandafter \@firstoftwo
 \else \expandafter \@secondoftwo
 \fi
}%
\providecommand \@ifx [1]{%
 \ifx #1\expandafter \@firstoftwo
 \else \expandafter \@secondoftwo
 \fi
}%
\providecommand \natexlab [1]{#1}%
\providecommand \enquote  [1]{``#1''}%
\providecommand \bibnamefont  [1]{#1}%
\providecommand \bibfnamefont [1]{#1}%
\providecommand \citenamefont [1]{#1}%
\providecommand \href@noop [0]{\@secondoftwo}%
\providecommand \href [0]{\begingroup \@sanitize@url \@href}%
\providecommand \@href[1]{\@@startlink{#1}\@@href}%
\providecommand \@@href[1]{\endgroup#1\@@endlink}%
\providecommand \@sanitize@url [0]{\catcode `\\12\catcode `\$12\catcode
  `\&12\catcode `\#12\catcode `\^12\catcode `\_12\catcode `\%12\relax}%
\providecommand \@@startlink[1]{}%
\providecommand \@@endlink[0]{}%
\providecommand \url  [0]{\begingroup\@sanitize@url \@url }%
\providecommand \@url [1]{\endgroup\@href {#1}{\urlprefix }}%
\providecommand \urlprefix  [0]{URL }%
\providecommand \Eprint [0]{\href }%
\providecommand \doibase [0]{https://doi.org/}%
\providecommand \selectlanguage [0]{\@gobble}%
\providecommand \bibinfo  [0]{\@secondoftwo}%
\providecommand \bibfield  [0]{\@secondoftwo}%
\providecommand \translation [1]{[#1]}%
\providecommand \BibitemOpen [0]{}%
\providecommand \bibitemStop [0]{}%
\providecommand \bibitemNoStop [0]{.\EOS\space}%
\providecommand \EOS [0]{\spacefactor3000\relax}%
\providecommand \BibitemShut  [1]{\csname bibitem#1\endcsname}%
\let\auto@bib@innerbib\@empty
\bibitem [{\citenamefont {Ahn}\ \emph {et~al.}(2019)\citenamefont {Ahn},
  \citenamefont {Park},\ and\ \citenamefont {Yang}}]{Ahn2019}%
  \BibitemOpen
  \bibfield  {author} {\bibinfo {author} {\bibfnamefont {J.}~\bibnamefont
  {Ahn}}, \bibinfo {author} {\bibfnamefont {S.}~\bibnamefont {Park}},\ and\
  \bibinfo {author} {\bibfnamefont {B.-J.}\ \bibnamefont {Yang}},\ }\bibfield
  {title} {\bibinfo {title} {Failure of {N}ielsen-{N}inomiya {T}heorem and
  {F}ragile {T}opology in {T}wo-{D}imensional {S}ystems with {S}pace-{T}ime
  {I}nversion {S}ymmetry: {A}pplication to {T}wisted {B}ilayer {G}raphene at
  {M}agic {A}ngle},\ }\href {https://doi.org/10.1103/PhysRevX.9.021013}
  {\bibfield  {journal} {\bibinfo  {journal} {Phys. Rev. X}\ }\textbf {\bibinfo
  {volume} {9}},\ \bibinfo {pages} {021013} (\bibinfo {year}
  {2019})}\BibitemShut {NoStop}%
\bibitem [{\citenamefont {Wu}\ \emph {et~al.}(2019)\citenamefont {Wu},
  \citenamefont {Soluyanov},\ and\ \citenamefont {Bzdusek}}]{Wu2019}%
  \BibitemOpen
  \bibfield  {author} {\bibinfo {author} {\bibfnamefont {Q.}~\bibnamefont
  {Wu}}, \bibinfo {author} {\bibfnamefont {A.~A.}\ \bibnamefont {Soluyanov}},\
  and\ \bibinfo {author} {\bibfnamefont {T.}~\bibnamefont {Bzdusek}},\
  }\bibfield  {title} {\bibinfo {title} {Non-{A}belian band topology in
  noninteracting metals},\ }\href {https://doi.org/10.1126/science.aau8740}
  {\bibfield  {journal} {\bibinfo  {journal} {Science}\ }\textbf {\bibinfo
  {volume} {365}},\ \bibinfo {pages} {1273} (\bibinfo {year}
  {2019})}\BibitemShut {NoStop}%
\bibitem [{\citenamefont {Bouhon}\ \emph
  {et~al.}(2020{\natexlab{a}})\citenamefont {Bouhon}, \citenamefont {Wu},
  \citenamefont {Slager}, \citenamefont {Weng}, \citenamefont {Yazyev},\ and\
  \citenamefont {Bzdu{\v s}ek}}]{Bouhon2019}%
  \BibitemOpen
  \bibfield  {author} {\bibinfo {author} {\bibfnamefont {A.}~\bibnamefont
  {Bouhon}}, \bibinfo {author} {\bibfnamefont {Q.}~\bibnamefont {Wu}}, \bibinfo
  {author} {\bibfnamefont {R.}~\bibnamefont {Slager}}, \bibinfo {author}
  {\bibfnamefont {H.}~\bibnamefont {Weng}}, \bibinfo {author} {\bibfnamefont
  {O.~V.}\ \bibnamefont {Yazyev}},\ and\ \bibinfo {author} {\bibfnamefont
  {T.}~\bibnamefont {Bzdu{\v s}ek}},\ }\bibfield  {title} {\bibinfo {title}
  {Non-{A}belian reciprocal braiding of {W}eyl points and its manifestation in
  {Z}r{T}e},\ }\href {https://doi.org/10.1038/s41567-020-0967-9} {\bibfield
  {journal} {\bibinfo  {journal} {Nature Physics}\ }\textbf {\bibinfo {volume}
  {16}},\ \bibinfo {pages} {1137} (\bibinfo {year}
  {2020}{\natexlab{a}})}\BibitemShut {NoStop}%
\bibitem [{\citenamefont {Bouhon}\ \emph
  {et~al.}(2020{\natexlab{b}})\citenamefont {Bouhon}, \citenamefont {Bzdusek},\
  and\ \citenamefont {Slager}}]{Bouhon2020}%
  \BibitemOpen
  \bibfield  {author} {\bibinfo {author} {\bibfnamefont {A.}~\bibnamefont
  {Bouhon}}, \bibinfo {author} {\bibfnamefont {T.}~\bibnamefont {Bzdusek}},\
  and\ \bibinfo {author} {\bibfnamefont {R.}~\bibnamefont {Slager}},\
  }\bibfield  {title} {\bibinfo {title} {Geometric approach to fragile topology
  beyond symmetry indicators},\ }\href
  {https://doi.org/10.1103/PhysRevB.102.115135} {\bibfield  {journal} {\bibinfo
   {journal} {Phys. Rev. B}\ }\textbf {\bibinfo {volume} {102}},\ \bibinfo
  {pages} {115135} (\bibinfo {year} {2020}{\natexlab{b}})}\BibitemShut
  {NoStop}%
\bibitem [{\citenamefont {\"Unal}\ \emph {et~al.}(2020)\citenamefont {\"Unal},
  \citenamefont {Bouhon},\ and\ \citenamefont {Slager}}]{Unal2020}%
  \BibitemOpen
  \bibfield  {author} {\bibinfo {author} {\bibfnamefont {F.~N.}\ \bibnamefont
  {\"Unal}}, \bibinfo {author} {\bibfnamefont {A.}~\bibnamefont {Bouhon}},\
  and\ \bibinfo {author} {\bibfnamefont {R.}~\bibnamefont {Slager}},\
  }\bibfield  {title} {\bibinfo {title} {Topological euler class as a dynamical
  observable in optical lattices},\ }\href
  {https://doi.org/10.1103/PhysRevLett.125.053601} {\bibfield  {journal}
  {\bibinfo  {journal} {Phys. Rev. Lett.}\ }\textbf {\bibinfo {volume} {125}},\
  \bibinfo {pages} {053601} (\bibinfo {year} {2020})}\BibitemShut {NoStop}%
\bibitem [{\citenamefont {Jiang}\ \emph {et~al.}(2021)\citenamefont {Jiang},
  \citenamefont {Bouhon}, \citenamefont {Lin}, \citenamefont {Zhou},
  \citenamefont {Hou}, \citenamefont {Li}, \citenamefont {Slager},\ and\
  \citenamefont {Jiang}}]{Jiang2021}%
  \BibitemOpen
  \bibfield  {author} {\bibinfo {author} {\bibfnamefont {B.}~\bibnamefont
  {Jiang}}, \bibinfo {author} {\bibfnamefont {A.}~\bibnamefont {Bouhon}},
  \bibinfo {author} {\bibfnamefont {Z.-K.}\ \bibnamefont {Lin}}, \bibinfo
  {author} {\bibfnamefont {X.}~\bibnamefont {Zhou}}, \bibinfo {author}
  {\bibfnamefont {B.}~\bibnamefont {Hou}}, \bibinfo {author} {\bibfnamefont
  {F.}~\bibnamefont {Li}}, \bibinfo {author} {\bibfnamefont {R.}~\bibnamefont
  {Slager}},\ and\ \bibinfo {author} {\bibfnamefont {J.-H.}\ \bibnamefont
  {Jiang}},\ }\bibfield  {title} {\bibinfo {title} {Experimental observation of
  non-{A}belian topological acoustic semimetals and their phase transitions},\
  }\href {https://doi.org/10.1038/s41567-021-01340-x} {\bibfield  {journal}
  {\bibinfo  {journal} {Nature Physics}\ }\textbf {\bibinfo {volume} {17}},\
  \bibinfo {pages} {1239–1246} (\bibinfo {year} {2021})}\BibitemShut
  {NoStop}%
\bibitem [{\citenamefont {Guo}\ \emph {et~al.}(2021)\citenamefont {Guo},
  \citenamefont {Jiang}, \citenamefont {Zhang}, \citenamefont {Zhang},
  \citenamefont {Zhang}, \citenamefont {Yang}, \citenamefont {Zhang},\ and\
  \citenamefont {Chan}}]{Guo2021}%
  \BibitemOpen
  \bibfield  {author} {\bibinfo {author} {\bibfnamefont {Q.}~\bibnamefont
  {Guo}}, \bibinfo {author} {\bibfnamefont {T.}~\bibnamefont {Jiang}}, \bibinfo
  {author} {\bibfnamefont {R.-Y.}\ \bibnamefont {Zhang}}, \bibinfo {author}
  {\bibfnamefont {L.}~\bibnamefont {Zhang}}, \bibinfo {author} {\bibfnamefont
  {Z.-Q.}\ \bibnamefont {Zhang}}, \bibinfo {author} {\bibfnamefont
  {B.}~\bibnamefont {Yang}}, \bibinfo {author} {\bibfnamefont {S.}~\bibnamefont
  {Zhang}},\ and\ \bibinfo {author} {\bibfnamefont {C.~T.}\ \bibnamefont
  {Chan}},\ }\bibfield  {title} {\bibinfo {title} {Experimental observation of
  non-abelian topological charges and edge states},\ }\href
  {https://doi.org/10.1038/s41586-021-03521-3} {\bibfield  {journal} {\bibinfo
  {journal} {Nature}\ }\textbf {\bibinfo {volume} {594}},\ \bibinfo {pages}
  {195} (\bibinfo {year} {2021})}\BibitemShut {NoStop}%
\bibitem [{\citenamefont {Zhao}\ \emph {et~al.}(2022)\citenamefont {Zhao},
  \citenamefont {Yang}, \citenamefont {Jiang}, \citenamefont {Mao},
  \citenamefont {Guo}, \citenamefont {Qiu}, \citenamefont {Wang}, \citenamefont
  {Yao}, \citenamefont {He}, \citenamefont {Zhou} \emph {et~al.}}]{Zhao2022}%
  \BibitemOpen
  \bibfield  {author} {\bibinfo {author} {\bibfnamefont {W.}~\bibnamefont
  {Zhao}}, \bibinfo {author} {\bibfnamefont {Y.-B.}\ \bibnamefont {Yang}},
  \bibinfo {author} {\bibfnamefont {Y.}~\bibnamefont {Jiang}}, \bibinfo
  {author} {\bibfnamefont {Z.}~\bibnamefont {Mao}}, \bibinfo {author}
  {\bibfnamefont {W.}~\bibnamefont {Guo}}, \bibinfo {author} {\bibfnamefont
  {L.}~\bibnamefont {Qiu}}, \bibinfo {author} {\bibfnamefont {G.}~\bibnamefont
  {Wang}}, \bibinfo {author} {\bibfnamefont {L.}~\bibnamefont {Yao}}, \bibinfo
  {author} {\bibfnamefont {L.}~\bibnamefont {He}}, \bibinfo {author}
  {\bibfnamefont {Z.}~\bibnamefont {Zhou}}, \emph {et~al.},\ }\bibfield
  {title} {\bibinfo {title} {Quantum simulation for topological euler
  insulators},\ }\href {https://doi.org/10.1038/s42005-022-01001-2} {\bibfield
  {journal} {\bibinfo  {journal} {Communications Physics}\ }\textbf {\bibinfo
  {volume} {5}},\ \bibinfo {pages} {223} (\bibinfo {year} {2022})}\BibitemShut
  {NoStop}%
\bibitem [{\citenamefont {Peng}\ \emph {et~al.}(2022)\citenamefont {Peng},
  \citenamefont {Bouhon}, \citenamefont {Monserrat},\ and\ \citenamefont
  {Slager}}]{Peng2022}%
  \BibitemOpen
  \bibfield  {author} {\bibinfo {author} {\bibfnamefont {B.}~\bibnamefont
  {Peng}}, \bibinfo {author} {\bibfnamefont {A.}~\bibnamefont {Bouhon}},
  \bibinfo {author} {\bibfnamefont {B.}~\bibnamefont {Monserrat}},\ and\
  \bibinfo {author} {\bibfnamefont {R.}~\bibnamefont {Slager}},\ }\bibfield
  {title} {\bibinfo {title} {Phonons as a platform for non-abelian braiding and
  its manifestation in layered silicates},\ }\href
  {https://doi.org/10.1038/s41467-022-28046-9} {\bibfield  {journal} {\bibinfo
  {journal} {Nature Communications}\ }\textbf {\bibinfo {volume} {13}},\
  \bibinfo {pages} {423} (\bibinfo {year} {2022})}\BibitemShut {NoStop}%
\bibitem [{\citenamefont {Yang}\ \emph {et~al.}(2024)\citenamefont {Yang},
  \citenamefont {Yang}, \citenamefont {Ma}, \citenamefont {Li}, \citenamefont
  {Zhang},\ and\ \citenamefont {Chan}}]{Yang2024}%
  \BibitemOpen
  \bibfield  {author} {\bibinfo {author} {\bibfnamefont {Y.}~\bibnamefont
  {Yang}}, \bibinfo {author} {\bibfnamefont {B.}~\bibnamefont {Yang}}, \bibinfo
  {author} {\bibfnamefont {G.}~\bibnamefont {Ma}}, \bibinfo {author}
  {\bibfnamefont {J.}~\bibnamefont {Li}}, \bibinfo {author} {\bibfnamefont
  {S.}~\bibnamefont {Zhang}},\ and\ \bibinfo {author} {\bibfnamefont {C.~T.}\
  \bibnamefont {Chan}},\ }\bibfield  {title} {\bibinfo {title} {Non-abelian
  physics in light and sound},\ }\href
  {https://doi.org/10.1126/science.adf9621} {\bibfield  {journal} {\bibinfo
  {journal} {Science}\ }\textbf {\bibinfo {volume} {383}},\ \bibinfo {pages}
  {eadf9621} (\bibinfo {year} {2024})}\BibitemShut {NoStop}%
\bibitem [{\citenamefont {Slager}\ \emph {et~al.}(2024)\citenamefont {Slager},
  \citenamefont {Bouhon},\ and\ \citenamefont {{\"U}nal}}]{Slager2024}%
  \BibitemOpen
  \bibfield  {author} {\bibinfo {author} {\bibfnamefont {R.}~\bibnamefont
  {Slager}}, \bibinfo {author} {\bibfnamefont {A.}~\bibnamefont {Bouhon}},\
  and\ \bibinfo {author} {\bibfnamefont {F.~N.}\ \bibnamefont {{\"U}nal}},\
  }\bibfield  {title} {\bibinfo {title} {Non-abelian floquet braiding and
  anomalous dirac string phase in periodically driven systems},\ }\href
  {https://doi.org/10.1038/s41467-024-45302-2} {\bibfield  {journal} {\bibinfo
  {journal} {Nature Communications}\ }\textbf {\bibinfo {volume} {15}},\
  \bibinfo {pages} {1144} (\bibinfo {year} {2024})}\BibitemShut {NoStop}%
\bibitem [{\citenamefont {Provost}\ and\ \citenamefont
  {Vallee}(1980)}]{Provost1980}%
  \BibitemOpen
  \bibfield  {author} {\bibinfo {author} {\bibfnamefont {J.}~\bibnamefont
  {Provost}}\ and\ \bibinfo {author} {\bibfnamefont {G.}~\bibnamefont
  {Vallee}},\ }\bibfield  {title} {\bibinfo {title} {Riemannian structure on
  manifolds of quantum states},\ }\href@noop {} {\bibfield  {journal} {\bibinfo
   {journal} {Communications in Mathematical Physics}\ }\textbf {\bibinfo
  {volume} {76}},\ \bibinfo {pages} {289} (\bibinfo {year} {1980})}\BibitemShut
  {NoStop}%
\bibitem [{\citenamefont {Resta}(2011)}]{Resta2011}%
  \BibitemOpen
  \bibfield  {author} {\bibinfo {author} {\bibfnamefont {R.}~\bibnamefont
  {Resta}},\ }\bibfield  {title} {\bibinfo {title} {The insulating state of
  matter: a geometrical theory},\ }\href
  {https://doi.org/10.1140/epjb/e2010-10874-4} {\bibfield  {journal} {\bibinfo
  {journal} {The European Physical Journal B}\ }\textbf {\bibinfo {volume}
  {79}},\ \bibinfo {pages} {121–137} (\bibinfo {year} {2011})}\BibitemShut
  {NoStop}%
\bibitem [{\citenamefont {Ma}\ \emph {et~al.}(2010)\citenamefont {Ma},
  \citenamefont {Chen}, \citenamefont {Fan},\ and\ \citenamefont
  {Liu}}]{Ma2010}%
  \BibitemOpen
  \bibfield  {author} {\bibinfo {author} {\bibfnamefont {Y.-Q.}\ \bibnamefont
  {Ma}}, \bibinfo {author} {\bibfnamefont {S.}~\bibnamefont {Chen}}, \bibinfo
  {author} {\bibfnamefont {H.}~\bibnamefont {Fan}},\ and\ \bibinfo {author}
  {\bibfnamefont {W.-M.}\ \bibnamefont {Liu}},\ }\bibfield  {title} {\bibinfo
  {title} {Abelian and non-abelian quantum geometric tensor},\ }\href
  {https://doi.org/10.1103/PhysRevB.81.245129} {\bibfield  {journal} {\bibinfo
  {journal} {Phys. Rev. B}\ }\textbf {\bibinfo {volume} {81}},\ \bibinfo
  {pages} {245129} (\bibinfo {year} {2010})}\BibitemShut {NoStop}%
\bibitem [{\citenamefont {Palumbo}(2021)}]{Palumbo2021}%
  \BibitemOpen
  \bibfield  {author} {\bibinfo {author} {\bibfnamefont {G.}~\bibnamefont
  {Palumbo}},\ }\bibfield  {title} {\bibinfo {title} {Non-abelian tensor berry
  connections in multiband topological systems},\ }\href
  {https://doi.org/10.1103/PhysRevLett.126.246801} {\bibfield  {journal}
  {\bibinfo  {journal} {Phys. Rev. Lett.}\ }\textbf {\bibinfo {volume} {126}},\
  \bibinfo {pages} {246801} (\bibinfo {year} {2021})}\BibitemShut {NoStop}%
\bibitem [{\citenamefont {Peotta}\ and\ \citenamefont
  {T{\"{o}}rm{\"{a}}}(2015)}]{Peotta2015}%
  \BibitemOpen
  \bibfield  {author} {\bibinfo {author} {\bibfnamefont {S.}~\bibnamefont
  {Peotta}}\ and\ \bibinfo {author} {\bibfnamefont {P.}~\bibnamefont
  {T{\"{o}}rm{\"{a}}}},\ }\bibfield  {title} {\bibinfo {title} {{Superfluidity
  in topologically nontrivial flat bands}},\ }\href
  {https://doi.org/10.1038/ncomms9944} {\bibfield  {journal} {\bibinfo
  {journal} {Nature Communications}\ }\textbf {\bibinfo {volume} {6}},\
  \bibinfo {pages} {8944} (\bibinfo {year} {2015})}\BibitemShut {NoStop}%
\bibitem [{\citenamefont {T{\"{o}}rm{\"{a}}}\ \emph {et~al.}(2022)\citenamefont
  {T{\"{o}}rm{\"{a}}}, \citenamefont {Peotta},\ and\ \citenamefont
  {Bernevig}}]{Torma2022}%
  \BibitemOpen
  \bibfield  {author} {\bibinfo {author} {\bibfnamefont {P.}~\bibnamefont
  {T{\"{o}}rm{\"{a}}}}, \bibinfo {author} {\bibfnamefont {S.}~\bibnamefont
  {Peotta}},\ and\ \bibinfo {author} {\bibfnamefont {B.~A.}\ \bibnamefont
  {Bernevig}},\ }\bibfield  {title} {\bibinfo {title} {{Superconductivity,
  superfluidity and quantum geometry in twisted multilayer systems}},\ }\href
  {https://doi.org/10.1038/s42254-022-00466-y} {\bibfield  {journal} {\bibinfo
  {journal} {Nature Reviews Physics}\ }\textbf {\bibinfo {volume} {4}},\
  \bibinfo {pages} {528} (\bibinfo {year} {2022})}\BibitemShut {NoStop}%
\bibitem [{\citenamefont {Gao}\ and\ \citenamefont {Xiao}(2019)}]{Gao2019}%
  \BibitemOpen
  \bibfield  {author} {\bibinfo {author} {\bibfnamefont {Y.}~\bibnamefont
  {Gao}}\ and\ \bibinfo {author} {\bibfnamefont {D.}~\bibnamefont {Xiao}},\
  }\bibfield  {title} {\bibinfo {title} {Nonreciprocal directional dichroism
  induced by the quantum metric dipole},\ }\href
  {https://doi.org/10.1103/PhysRevLett.122.227402} {\bibfield  {journal}
  {\bibinfo  {journal} {Phys. Rev. Lett.}\ }\textbf {\bibinfo {volume} {122}},\
  \bibinfo {pages} {227402} (\bibinfo {year} {2019})}\BibitemShut {NoStop}%
\bibitem [{\citenamefont {Ahn}\ \emph {et~al.}(2020)\citenamefont {Ahn},
  \citenamefont {Guo},\ and\ \citenamefont {Nagaosa}}]{Ahn2020}%
  \BibitemOpen
  \bibfield  {author} {\bibinfo {author} {\bibfnamefont {J.}~\bibnamefont
  {Ahn}}, \bibinfo {author} {\bibfnamefont {G.-Y.}\ \bibnamefont {Guo}},\ and\
  \bibinfo {author} {\bibfnamefont {N.}~\bibnamefont {Nagaosa}},\ }\bibfield
  {title} {\bibinfo {title} {Low-frequency divergence and quantum geometry of
  the bulk photovoltaic effect in topological semimetals},\ }\href
  {https://doi.org/10.1103/PhysRevX.10.041041} {\bibfield  {journal} {\bibinfo
  {journal} {Phys. Rev. X}\ }\textbf {\bibinfo {volume} {10}},\ \bibinfo
  {pages} {041041} (\bibinfo {year} {2020})}\BibitemShut {NoStop}%
\bibitem [{\citenamefont {Bouhon}\ \emph {et~al.}(2023)\citenamefont {Bouhon},
  \citenamefont {Timmel},\ and\ \citenamefont {Slager}}]{Bouhon2023}%
  \BibitemOpen
  \bibfield  {author} {\bibinfo {author} {\bibfnamefont {A.}~\bibnamefont
  {Bouhon}}, \bibinfo {author} {\bibfnamefont {A.}~\bibnamefont {Timmel}},\
  and\ \bibinfo {author} {\bibfnamefont {R.}~\bibnamefont {Slager}},\ }\href
  {https://arxiv.org/abs/2303.02180} {\bibinfo {title} {Quantum geometry beyond
  projective single bands}} (\bibinfo {year} {2023}),\ \Eprint
  {https://arxiv.org/abs/2303.02180} {arXiv:2303.02180 [cond-mat.mes-hall]}
  \BibitemShut {NoStop}%
\bibitem [{\citenamefont {Jankowski}\ \emph {et~al.}(2025)\citenamefont
  {Jankowski}, \citenamefont {Morris}, \citenamefont {Bouhon}, \citenamefont
  {\"Unal},\ and\ \citenamefont {Slager}}]{Jankowski2023}%
  \BibitemOpen
  \bibfield  {author} {\bibinfo {author} {\bibfnamefont {W.~J.}\ \bibnamefont
  {Jankowski}}, \bibinfo {author} {\bibfnamefont {A.~S.}\ \bibnamefont
  {Morris}}, \bibinfo {author} {\bibfnamefont {A.}~\bibnamefont {Bouhon}},
  \bibinfo {author} {\bibfnamefont {F.~N.}\ \bibnamefont {\"Unal}},\ and\
  \bibinfo {author} {\bibfnamefont {R.}~\bibnamefont {Slager}},\ }\bibfield
  {title} {\bibinfo {title} {Optical manifestations and bounds of topological
  euler class},\ }\href {https://doi.org/10.1103/PhysRevB.111.L081103}
  {\bibfield  {journal} {\bibinfo  {journal} {Phys. Rev. B}\ }\textbf {\bibinfo
  {volume} {111}},\ \bibinfo {pages} {L081103} (\bibinfo {year}
  {2025})}\BibitemShut {NoStop}%
\bibitem [{\citenamefont {Tanaka}\ \emph {et~al.}(2024)\citenamefont {Tanaka},
  \citenamefont {Watanabe},\ and\ \citenamefont {Yanase}}]{Tanaka2024}%
  \BibitemOpen
  \bibfield  {author} {\bibinfo {author} {\bibfnamefont {H.}~\bibnamefont
  {Tanaka}}, \bibinfo {author} {\bibfnamefont {H.}~\bibnamefont {Watanabe}},\
  and\ \bibinfo {author} {\bibfnamefont {Y.}~\bibnamefont {Yanase}},\
  }\bibfield  {title} {\bibinfo {title} {Nonlinear optical response in
  superconductors in magnetic field: Quantum geometry and topological
  superconductivity},\ }\href {https://doi.org/10.1103/PhysRevB.110.014520}
  {\bibfield  {journal} {\bibinfo  {journal} {Phys. Rev. B}\ }\textbf {\bibinfo
  {volume} {110}},\ \bibinfo {pages} {014520} (\bibinfo {year}
  {2024})}\BibitemShut {NoStop}%
\bibitem [{\citenamefont {Fang}\ \emph {et~al.}(2024)\citenamefont {Fang},
  \citenamefont {Cano},\ and\ \citenamefont {Ghorashi}}]{Fang2023}%
  \BibitemOpen
  \bibfield  {author} {\bibinfo {author} {\bibfnamefont {Y.}~\bibnamefont
  {Fang}}, \bibinfo {author} {\bibfnamefont {J.}~\bibnamefont {Cano}},\ and\
  \bibinfo {author} {\bibfnamefont {S.~A.~A.}\ \bibnamefont {Ghorashi}},\
  }\bibfield  {title} {\bibinfo {title} {Quantum geometry induced nonlinear
  transport in altermagnets},\ }\href
  {https://doi.org/10.1103/PhysRevLett.133.106701} {\bibfield  {journal}
  {\bibinfo  {journal} {Phys. Rev. Lett.}\ }\textbf {\bibinfo {volume} {133}},\
  \bibinfo {pages} {106701} (\bibinfo {year} {2024})}\BibitemShut {NoStop}%
\bibitem [{\citenamefont {Li}\ \emph {et~al.}(2024{\natexlab{a}})\citenamefont
  {Li}, \citenamefont {Zhang}, \citenamefont {Zhou}, \citenamefont {Ma},
  \citenamefont {Lei}, \citenamefont {Jin}, \citenamefont {He}, \citenamefont
  {Li}, \citenamefont {Law},\ and\ \citenamefont {Wang}}]{Li2024b}%
  \BibitemOpen
  \bibfield  {author} {\bibinfo {author} {\bibfnamefont {H.}~\bibnamefont
  {Li}}, \bibinfo {author} {\bibfnamefont {C.}~\bibnamefont {Zhang}}, \bibinfo
  {author} {\bibfnamefont {C.}~\bibnamefont {Zhou}}, \bibinfo {author}
  {\bibfnamefont {C.}~\bibnamefont {Ma}}, \bibinfo {author} {\bibfnamefont
  {X.}~\bibnamefont {Lei}}, \bibinfo {author} {\bibfnamefont {Z.}~\bibnamefont
  {Jin}}, \bibinfo {author} {\bibfnamefont {H.}~\bibnamefont {He}}, \bibinfo
  {author} {\bibfnamefont {B.}~\bibnamefont {Li}}, \bibinfo {author}
  {\bibfnamefont {K.~T.}\ \bibnamefont {Law}},\ and\ \bibinfo {author}
  {\bibfnamefont {J.}~\bibnamefont {Wang}},\ }\bibfield  {title} {\bibinfo
  {title} {Quantum geometry quadrupole-induced third-order nonlinear transport
  in antiferromagnetic topological insulator {M}n{B}i2{T}e4},\ }\href
  {https://doi.org/10.1038/s41467-024-52206-8} {\bibfield  {journal} {\bibinfo
  {journal} {Nature Communications}\ }\textbf {\bibinfo {volume} {15}},\
  \bibinfo {pages} {7779} (\bibinfo {year} {2024}{\natexlab{a}})}\BibitemShut
  {NoStop}%
\bibitem [{\citenamefont {Jankowski}\ and\ \citenamefont
  {Slager}(2024)}]{Jankowski2024a}%
  \BibitemOpen
  \bibfield  {author} {\bibinfo {author} {\bibfnamefont {W.~J.}\ \bibnamefont
  {Jankowski}}\ and\ \bibinfo {author} {\bibfnamefont {R.}~\bibnamefont
  {Slager}},\ }\bibfield  {title} {\bibinfo {title} {Quantized integrated shift
  effect in multigap topological phases},\ }\href
  {https://doi.org/10.1103/PhysRevLett.133.186601} {\bibfield  {journal}
  {\bibinfo  {journal} {Phys. Rev. Lett.}\ }\textbf {\bibinfo {volume} {133}},\
  \bibinfo {pages} {186601} (\bibinfo {year} {2024})}\BibitemShut {NoStop}%
\bibitem [{\citenamefont {Mandal}\ \emph {et~al.}(2024)\citenamefont {Mandal},
  \citenamefont {Sarkar}, \citenamefont {Das},\ and\ \citenamefont
  {Agarwal}}]{Mandal2024}%
  \BibitemOpen
  \bibfield  {author} {\bibinfo {author} {\bibfnamefont {D.}~\bibnamefont
  {Mandal}}, \bibinfo {author} {\bibfnamefont {S.}~\bibnamefont {Sarkar}},
  \bibinfo {author} {\bibfnamefont {K.}~\bibnamefont {Das}},\ and\ \bibinfo
  {author} {\bibfnamefont {A.}~\bibnamefont {Agarwal}},\ }\bibfield  {title}
  {\bibinfo {title} {Quantum geometry induced third-order nonlinear transport
  responses},\ }\href {https://doi.org/10.1103/PhysRevB.110.195131} {\bibfield
  {journal} {\bibinfo  {journal} {Phys. Rev. B}\ }\textbf {\bibinfo {volume}
  {110}},\ \bibinfo {pages} {195131} (\bibinfo {year} {2024})}\BibitemShut
  {NoStop}%
\bibitem [{\citenamefont {Mercaldo}\ \emph {et~al.}(2025)\citenamefont
  {Mercaldo}, \citenamefont {Cuoco},\ and\ \citenamefont
  {Ortix}}]{Mercaldo2025}%
  \BibitemOpen
  \bibfield  {author} {\bibinfo {author} {\bibfnamefont {M.~T.}\ \bibnamefont
  {Mercaldo}}, \bibinfo {author} {\bibfnamefont {M.}~\bibnamefont {Cuoco}},\
  and\ \bibinfo {author} {\bibfnamefont {C.}~\bibnamefont {Ortix}},\ }\bibfield
   {title} {\bibinfo {title} {Nonlinear planar magnetotransport as a probe of
  the topology of surface states},\ }\href
  {https://doi.org/10.1103/PhysRevB.111.155442} {\bibfield  {journal} {\bibinfo
   {journal} {Phys. Rev. B}\ }\textbf {\bibinfo {volume} {111}},\ \bibinfo
  {pages} {155442} (\bibinfo {year} {2025})}\BibitemShut {NoStop}%
\bibitem [{\citenamefont {Yu}\ \emph {et~al.}(2024)\citenamefont {Yu},
  \citenamefont {Li}, \citenamefont {Chu}, \citenamefont {Mera}, \citenamefont
  {Ünal}, \citenamefont {Yang}, \citenamefont {Liu}, \citenamefont {Goldman},\
  and\ \citenamefont {Cai}}]{Yu2024}%
  \BibitemOpen
  \bibfield  {author} {\bibinfo {author} {\bibfnamefont {M.}~\bibnamefont
  {Yu}}, \bibinfo {author} {\bibfnamefont {X.}~\bibnamefont {Li}}, \bibinfo
  {author} {\bibfnamefont {Y.}~\bibnamefont {Chu}}, \bibinfo {author}
  {\bibfnamefont {B.}~\bibnamefont {Mera}}, \bibinfo {author} {\bibfnamefont
  {F.~N.}\ \bibnamefont {Ünal}}, \bibinfo {author} {\bibfnamefont
  {P.}~\bibnamefont {Yang}}, \bibinfo {author} {\bibfnamefont {Y.}~\bibnamefont
  {Liu}}, \bibinfo {author} {\bibfnamefont {N.}~\bibnamefont {Goldman}},\ and\
  \bibinfo {author} {\bibfnamefont {J.}~\bibnamefont {Cai}},\ }\bibfield
  {title} {\bibinfo {title} {Experimental demonstration of topological bounds
  in quantum metrology},\ }\href {https://doi.org/10.1093/nsr/nwae065}
  {\bibfield  {journal} {\bibinfo  {journal} {National Science Review}\
  }\textbf {\bibinfo {volume} {11}},\ \bibinfo {pages} {nwae065} (\bibinfo
  {year} {2024})}\BibitemShut {NoStop}%
\bibitem [{\citenamefont {Li}\ \emph {et~al.}(2024{\natexlab{b}})\citenamefont
  {Li}, \citenamefont {Chen},\ and\ \citenamefont {Cappellaro}}]{Li2024a}%
  \BibitemOpen
  \bibfield  {author} {\bibinfo {author} {\bibfnamefont {C.}~\bibnamefont
  {Li}}, \bibinfo {author} {\bibfnamefont {M.}~\bibnamefont {Chen}},\ and\
  \bibinfo {author} {\bibfnamefont {P.}~\bibnamefont {Cappellaro}},\ }\href
  {https://arxiv.org/abs/2204.13777} {\bibinfo {title} {A geometric
  perspective: experimental evaluation of the quantum cramer-rao bound}}
  (\bibinfo {year} {2024}{\natexlab{b}}),\ \Eprint
  {https://arxiv.org/abs/2204.13777} {arXiv:2204.13777 [quant-ph]} \BibitemShut
  {NoStop}%
\bibitem [{\citenamefont {Rhim}\ \emph {et~al.}(2020)\citenamefont {Rhim},
  \citenamefont {Kim},\ and\ \citenamefont {Yang}}]{Rhim2020}%
  \BibitemOpen
  \bibfield  {author} {\bibinfo {author} {\bibfnamefont {J.-W.}\ \bibnamefont
  {Rhim}}, \bibinfo {author} {\bibfnamefont {K.}~\bibnamefont {Kim}},\ and\
  \bibinfo {author} {\bibfnamefont {B.-J.}\ \bibnamefont {Yang}},\ }\bibfield
  {title} {\bibinfo {title} {{Quantum distance and anomalous Landau levels of
  flat bands}},\ }\href {https://doi.org/10.1038/s41586-020-2540-1} {\bibfield
  {journal} {\bibinfo  {journal} {Nature}\ }\textbf {\bibinfo {volume} {584}},\
  \bibinfo {pages} {59} (\bibinfo {year} {2020})}\BibitemShut {NoStop}%
\bibitem [{\citenamefont {Lai}\ \emph {et~al.}(2021)\citenamefont {Lai},
  \citenamefont {Liu}, \citenamefont {Zhang}, \citenamefont {Zhao},
  \citenamefont {Feng}, \citenamefont {Wang}, \citenamefont {Tang},
  \citenamefont {Liu}, \citenamefont {Novoselov}, \citenamefont {Yang},\ and\
  \citenamefont {Gao}}]{Lai2021}%
  \BibitemOpen
  \bibfield  {author} {\bibinfo {author} {\bibfnamefont {S.}~\bibnamefont
  {Lai}}, \bibinfo {author} {\bibfnamefont {H.}~\bibnamefont {Liu}}, \bibinfo
  {author} {\bibfnamefont {Z.}~\bibnamefont {Zhang}}, \bibinfo {author}
  {\bibfnamefont {J.}~\bibnamefont {Zhao}}, \bibinfo {author} {\bibfnamefont
  {X.}~\bibnamefont {Feng}}, \bibinfo {author} {\bibfnamefont {N.}~\bibnamefont
  {Wang}}, \bibinfo {author} {\bibfnamefont {C.}~\bibnamefont {Tang}}, \bibinfo
  {author} {\bibfnamefont {Y.}~\bibnamefont {Liu}}, \bibinfo {author}
  {\bibfnamefont {K.~S.}\ \bibnamefont {Novoselov}}, \bibinfo {author}
  {\bibfnamefont {S.~A.}\ \bibnamefont {Yang}},\ and\ \bibinfo {author}
  {\bibfnamefont {W.-b.}\ \bibnamefont {Gao}},\ }\bibfield  {title} {\bibinfo
  {title} {Third-order nonlinear {H}all effect induced by the
  {B}erry-connection polarizability tensor},\ }\href
  {https://doi.org/10.1038/s41565-021-00917-0} {\bibfield  {journal} {\bibinfo
  {journal} {Nature Nanotechnology}\ }\textbf {\bibinfo {volume} {16}},\
  \bibinfo {pages} {869} (\bibinfo {year} {2021})}\BibitemShut {NoStop}%
\bibitem [{\citenamefont {Gao}\ \emph {et~al.}(2023)\citenamefont {Gao},
  \citenamefont {Liu}, \citenamefont {Qiu}, \citenamefont {Ghosh},
  \citenamefont {V.~Trevisan}, \citenamefont {Onishi}, \citenamefont {Hu},
  \citenamefont {Qian}, \citenamefont {Tien}, \citenamefont {Chen},
  \citenamefont {Huang}, \citenamefont {Bérubé}, \citenamefont {Li},
  \citenamefont {Tzschaschel}, \citenamefont {Dinh}, \citenamefont {Sun},
  \citenamefont {Ho}, \citenamefont {Lien}, \citenamefont {Singh},
  \citenamefont {Watanabe}, \citenamefont {Taniguchi}, \citenamefont {Bell},
  \citenamefont {Lin}, \citenamefont {Chang}, \citenamefont {Du}, \citenamefont
  {Bansil}, \citenamefont {Fu}, \citenamefont {Ni}, \citenamefont {Orth},
  \citenamefont {Ma},\ and\ \citenamefont {Xu}}]{Gao2023a}%
  \BibitemOpen
  \bibfield  {author} {\bibinfo {author} {\bibfnamefont {A.}~\bibnamefont
  {Gao}}, \bibinfo {author} {\bibfnamefont {Y.-F.}\ \bibnamefont {Liu}},
  \bibinfo {author} {\bibfnamefont {J.-X.}\ \bibnamefont {Qiu}}, \bibinfo
  {author} {\bibfnamefont {B.}~\bibnamefont {Ghosh}}, \bibinfo {author}
  {\bibfnamefont {T.}~\bibnamefont {V.~Trevisan}}, \bibinfo {author}
  {\bibfnamefont {Y.}~\bibnamefont {Onishi}}, \bibinfo {author} {\bibfnamefont
  {C.}~\bibnamefont {Hu}}, \bibinfo {author} {\bibfnamefont {T.}~\bibnamefont
  {Qian}}, \bibinfo {author} {\bibfnamefont {H.-J.}\ \bibnamefont {Tien}},
  \bibinfo {author} {\bibfnamefont {S.-W.}\ \bibnamefont {Chen}}, \bibinfo
  {author} {\bibfnamefont {M.}~\bibnamefont {Huang}}, \bibinfo {author}
  {\bibfnamefont {D.}~\bibnamefont {Bérubé}}, \bibinfo {author}
  {\bibfnamefont {H.}~\bibnamefont {Li}}, \bibinfo {author} {\bibfnamefont
  {C.}~\bibnamefont {Tzschaschel}}, \bibinfo {author} {\bibfnamefont
  {T.}~\bibnamefont {Dinh}}, \bibinfo {author} {\bibfnamefont {Z.}~\bibnamefont
  {Sun}}, \bibinfo {author} {\bibfnamefont {S.-C.}\ \bibnamefont {Ho}},
  \bibinfo {author} {\bibfnamefont {S.-W.}\ \bibnamefont {Lien}}, \bibinfo
  {author} {\bibfnamefont {B.}~\bibnamefont {Singh}}, \bibinfo {author}
  {\bibfnamefont {K.}~\bibnamefont {Watanabe}}, \bibinfo {author}
  {\bibfnamefont {T.}~\bibnamefont {Taniguchi}}, \bibinfo {author}
  {\bibfnamefont {D.~C.}\ \bibnamefont {Bell}}, \bibinfo {author}
  {\bibfnamefont {H.}~\bibnamefont {Lin}}, \bibinfo {author} {\bibfnamefont
  {T.-R.}\ \bibnamefont {Chang}}, \bibinfo {author} {\bibfnamefont {C.~R.}\
  \bibnamefont {Du}}, \bibinfo {author} {\bibfnamefont {A.}~\bibnamefont
  {Bansil}}, \bibinfo {author} {\bibfnamefont {L.}~\bibnamefont {Fu}}, \bibinfo
  {author} {\bibfnamefont {N.}~\bibnamefont {Ni}}, \bibinfo {author}
  {\bibfnamefont {P.~P.}\ \bibnamefont {Orth}}, \bibinfo {author}
  {\bibfnamefont {Q.}~\bibnamefont {Ma}},\ and\ \bibinfo {author}
  {\bibfnamefont {S.-Y.}\ \bibnamefont {Xu}},\ }\bibfield  {title} {\bibinfo
  {title} {Quantum metric nonlinear {H}all effect in a topological
  antiferromagnetic heterostructure},\ }\href
  {https://doi.org/10.1126/science.adf1506} {\bibfield  {journal} {\bibinfo
  {journal} {Science}\ }\textbf {\bibinfo {volume} {381}},\ \bibinfo {pages}
  {181–186} (\bibinfo {year} {2023})}\BibitemShut {NoStop}%
\bibitem [{\citenamefont {Hasan}\ and\ \citenamefont {Kane}(2010)}]{Hasan2010}%
  \BibitemOpen
  \bibfield  {author} {\bibinfo {author} {\bibfnamefont {M.~Z.}\ \bibnamefont
  {Hasan}}\ and\ \bibinfo {author} {\bibfnamefont {C.~L.}\ \bibnamefont
  {Kane}},\ }\bibfield  {title} {\bibinfo {title} {Colloquium: Topological
  insulators},\ }\href {https://doi.org/10.1103/RevModPhys.82.3045} {\bibfield
  {journal} {\bibinfo  {journal} {Rev. Mod. Phys.}\ }\textbf {\bibinfo {volume}
  {82}},\ \bibinfo {pages} {3045} (\bibinfo {year} {2010})}\BibitemShut
  {NoStop}%
\bibitem [{\citenamefont {Chiu}\ \emph {et~al.}(2016)\citenamefont {Chiu},
  \citenamefont {Teo}, \citenamefont {Schnyder},\ and\ \citenamefont
  {Ryu}}]{Chiu2016}%
  \BibitemOpen
  \bibfield  {author} {\bibinfo {author} {\bibfnamefont {C.-K.}\ \bibnamefont
  {Chiu}}, \bibinfo {author} {\bibfnamefont {J.~C.~Y.}\ \bibnamefont {Teo}},
  \bibinfo {author} {\bibfnamefont {A.~P.}\ \bibnamefont {Schnyder}},\ and\
  \bibinfo {author} {\bibfnamefont {S.}~\bibnamefont {Ryu}},\ }\bibfield
  {title} {\bibinfo {title} {Classification of topological quantum matter with
  symmetries},\ }\href {https://doi.org/10.1103/RevModPhys.88.035005}
  {\bibfield  {journal} {\bibinfo  {journal} {Rev. Mod. Phys.}\ }\textbf
  {\bibinfo {volume} {88}},\ \bibinfo {pages} {035005} (\bibinfo {year}
  {2016})}\BibitemShut {NoStop}%
\bibitem [{\citenamefont {Bouhon}\ \emph {et~al.}(2019)\citenamefont {Bouhon},
  \citenamefont {{Black-Schaffer}},\ and\ \citenamefont {Slager}}]{Bouhon2018}%
  \BibitemOpen
  \bibfield  {author} {\bibinfo {author} {\bibfnamefont {A.}~\bibnamefont
  {Bouhon}}, \bibinfo {author} {\bibfnamefont {A.~M.}\ \bibnamefont
  {{Black-Schaffer}}},\ and\ \bibinfo {author} {\bibfnamefont {R.}~\bibnamefont
  {Slager}},\ }\bibfield  {title} {\bibinfo {title} {Wilson loop approach to
  fragile topology of split elementary band representations and topological
  crystalline insulators with time-reversal symmetry},\ }\href
  {https://doi.org/10.1103/PhysRevB.100.195135} {\bibfield  {journal} {\bibinfo
   {journal} {Phys. Rev. B}\ }\textbf {\bibinfo {volume} {100}},\ \bibinfo
  {pages} {195135} (\bibinfo {year} {2019})}\BibitemShut {NoStop}%
\bibitem [{\citenamefont {Yang}\ \emph {et~al.}(2020)\citenamefont {Yang},
  \citenamefont {Yang}, \citenamefont {You}, \citenamefont {Chan},
  \citenamefont {Mao}, \citenamefont {Guo}, \citenamefont {Ma}, \citenamefont
  {Xia}, \citenamefont {Fan}, \citenamefont {Xiang},\ and\ \citenamefont
  {Zhang}}]{Yang2020}%
  \BibitemOpen
  \bibfield  {author} {\bibinfo {author} {\bibfnamefont {E.}~\bibnamefont
  {Yang}}, \bibinfo {author} {\bibfnamefont {B.}~\bibnamefont {Yang}}, \bibinfo
  {author} {\bibfnamefont {O.}~\bibnamefont {You}}, \bibinfo {author}
  {\bibfnamefont {H.-C.}\ \bibnamefont {Chan}}, \bibinfo {author}
  {\bibfnamefont {P.}~\bibnamefont {Mao}}, \bibinfo {author} {\bibfnamefont
  {Q.}~\bibnamefont {Guo}}, \bibinfo {author} {\bibfnamefont {S.}~\bibnamefont
  {Ma}}, \bibinfo {author} {\bibfnamefont {L.}~\bibnamefont {Xia}}, \bibinfo
  {author} {\bibfnamefont {D.}~\bibnamefont {Fan}}, \bibinfo {author}
  {\bibfnamefont {Y.}~\bibnamefont {Xiang}},\ and\ \bibinfo {author}
  {\bibfnamefont {S.}~\bibnamefont {Zhang}},\ }\bibfield  {title} {\bibinfo
  {title} {Observation of non-abelian nodal links in photonics},\ }\href
  {https://doi.org/10.1103/PhysRevLett.125.033901} {\bibfield  {journal}
  {\bibinfo  {journal} {Phys. Rev. Lett.}\ }\textbf {\bibinfo {volume} {125}},\
  \bibinfo {pages} {033901} (\bibinfo {year} {2020})}\BibitemShut {NoStop}%
\bibitem [{\citenamefont {Wang}\ \emph {et~al.}(2021)\citenamefont {Wang},
  \citenamefont {Dutt}, \citenamefont {Wojcik},\ and\ \citenamefont
  {Fan}}]{Wang2021}%
  \BibitemOpen
  \bibfield  {author} {\bibinfo {author} {\bibfnamefont {K.}~\bibnamefont
  {Wang}}, \bibinfo {author} {\bibfnamefont {A.}~\bibnamefont {Dutt}}, \bibinfo
  {author} {\bibfnamefont {C.~C.}\ \bibnamefont {Wojcik}},\ and\ \bibinfo
  {author} {\bibfnamefont {S.}~\bibnamefont {Fan}},\ }\bibfield  {title}
  {\bibinfo {title} {{Topological complex-energy braiding of non-Hermitian
  bands}},\ }\href {https://doi.org/10.1038/s41586-021-03848-x} {\bibfield
  {journal} {\bibinfo  {journal} {Nature}\ }\textbf {\bibinfo {volume} {598}},\
  \bibinfo {pages} {59} (\bibinfo {year} {2021})}\BibitemShut {NoStop}%
\bibitem [{\citenamefont {Breach}\ \emph {et~al.}(2024)\citenamefont {Breach},
  \citenamefont {Slager},\ and\ \citenamefont {\"Unal}}]{Breach2024}%
  \BibitemOpen
  \bibfield  {author} {\bibinfo {author} {\bibfnamefont {O.}~\bibnamefont
  {Breach}}, \bibinfo {author} {\bibfnamefont {R.-J.}\ \bibnamefont {Slager}},\
  and\ \bibinfo {author} {\bibfnamefont {F.~N.}\ \bibnamefont {\"Unal}},\
  }\bibfield  {title} {\bibinfo {title} {Interferometry of non-abelian band
  singularities and euler class topology},\ }\href
  {https://doi.org/10.1103/PhysRevLett.133.093404} {\bibfield  {journal}
  {\bibinfo  {journal} {Phys. Rev. Lett.}\ }\textbf {\bibinfo {volume} {133}},\
  \bibinfo {pages} {093404} (\bibinfo {year} {2024})}\BibitemShut {NoStop}%
\bibitem [{\citenamefont {Kobayashi}\ \emph {et~al.}(2025)\citenamefont
  {Kobayashi}, \citenamefont {Sato},\ and\ \citenamefont
  {Furusaki}}]{Kobayashi2025}%
  \BibitemOpen
  \bibfield  {author} {\bibinfo {author} {\bibfnamefont {S.}~\bibnamefont
  {Kobayashi}}, \bibinfo {author} {\bibfnamefont {M.}~\bibnamefont {Sato}},\
  and\ \bibinfo {author} {\bibfnamefont {A.}~\bibnamefont {Furusaki}},\
  }\bibfield  {title} {\bibinfo {title} {Euler band topology in superfluids and
  superconductors},\ }\href@noop {} {\bibfield  {journal} {\bibinfo  {journal}
  {arXiv preprint arXiv:2509.06406}\ } (\bibinfo {year} {2025})}\BibitemShut
  {NoStop}%
\bibitem [{\citenamefont {Wahl}\ \emph {et~al.}(2025)\citenamefont {Wahl},
  \citenamefont {Jankowski}, \citenamefont {Bouhon}, \citenamefont
  {Chaudhary},\ and\ \citenamefont {Slager}}]{Wahl2024}%
  \BibitemOpen
  \bibfield  {author} {\bibinfo {author} {\bibfnamefont {T.~B.}\ \bibnamefont
  {Wahl}}, \bibinfo {author} {\bibfnamefont {W.~J.}\ \bibnamefont {Jankowski}},
  \bibinfo {author} {\bibfnamefont {A.}~\bibnamefont {Bouhon}}, \bibinfo
  {author} {\bibfnamefont {G.}~\bibnamefont {Chaudhary}},\ and\ \bibinfo
  {author} {\bibfnamefont {R.}~\bibnamefont {Slager}},\ }\bibfield  {title}
  {\bibinfo {title} {Exact projected entangled pair ground states with
  topological euler invariant},\ }\href
  {https://doi.org/10.1038/s41467-024-55484-4} {\bibfield  {journal} {\bibinfo
  {journal} {Nature Communications}\ }\textbf {\bibinfo {volume} {16}},\
  \bibinfo {pages} {284} (\bibinfo {year} {2025})}\BibitemShut {NoStop}%
\bibitem [{\citenamefont {Hu}\ \emph {et~al.}(2024)\citenamefont {Hu},
  \citenamefont {Tong}, \citenamefont {Jiang}, \citenamefont {Jiang},
  \citenamefont {Chen},\ and\ \citenamefont {Yang}}]{Hu2024}%
  \BibitemOpen
  \bibfield  {author} {\bibinfo {author} {\bibfnamefont {Y.}~\bibnamefont
  {Hu}}, \bibinfo {author} {\bibfnamefont {M.}~\bibnamefont {Tong}}, \bibinfo
  {author} {\bibfnamefont {T.}~\bibnamefont {Jiang}}, \bibinfo {author}
  {\bibfnamefont {J.-H.}\ \bibnamefont {Jiang}}, \bibinfo {author}
  {\bibfnamefont {H.}~\bibnamefont {Chen}},\ and\ \bibinfo {author}
  {\bibfnamefont {Y.}~\bibnamefont {Yang}},\ }\bibfield  {title} {\bibinfo
  {title} {Observation of two-dimensional time-reversal broken non-abelian
  topological states},\ }\href {https://doi.org/10.1038/s41467-024-54403-x}
  {\bibfield  {journal} {\bibinfo  {journal} {Nature Communications}\ }\textbf
  {\bibinfo {volume} {15}},\ \bibinfo {pages} {10036} (\bibinfo {year}
  {2024})}\BibitemShut {NoStop}%
\bibitem [{\citenamefont {Li}\ \emph {et~al.}(2016)\citenamefont {Li},
  \citenamefont {Duca}, \citenamefont {Reitter}, \citenamefont {Grusdt},
  \citenamefont {Demler}, \citenamefont {Endres}, \citenamefont
  {Schleier-Smith}, \citenamefont {Bloch},\ and\ \citenamefont
  {Schneider}}]{Li2016}%
  \BibitemOpen
  \bibfield  {author} {\bibinfo {author} {\bibfnamefont {T.}~\bibnamefont
  {Li}}, \bibinfo {author} {\bibfnamefont {L.}~\bibnamefont {Duca}}, \bibinfo
  {author} {\bibfnamefont {M.}~\bibnamefont {Reitter}}, \bibinfo {author}
  {\bibfnamefont {F.}~\bibnamefont {Grusdt}}, \bibinfo {author} {\bibfnamefont
  {E.}~\bibnamefont {Demler}}, \bibinfo {author} {\bibfnamefont
  {M.}~\bibnamefont {Endres}}, \bibinfo {author} {\bibfnamefont
  {M.}~\bibnamefont {Schleier-Smith}}, \bibinfo {author} {\bibfnamefont
  {I.}~\bibnamefont {Bloch}},\ and\ \bibinfo {author} {\bibfnamefont
  {U.}~\bibnamefont {Schneider}},\ }\bibfield  {title} {\bibinfo {title} {Bloch
  state tomography using wilson lines},\ }\href
  {https://doi.org/10.1126/science.aad5812} {\bibfield  {journal} {\bibinfo
  {journal} {Science}\ }\textbf {\bibinfo {volume} {352}},\ \bibinfo {pages}
  {1094} (\bibinfo {year} {2016})}\BibitemShut {NoStop}%
\bibitem [{\citenamefont {Fl{\"a}schner}\ \emph {et~al.}(2016)\citenamefont
  {Fl{\"a}schner}, \citenamefont {Rem}, \citenamefont {Tarnowski},
  \citenamefont {Vogel}, \citenamefont {Lühmann}, \citenamefont {Sengstock},\
  and\ \citenamefont {Weitenberg}}]{Flaschner2016}%
  \BibitemOpen
  \bibfield  {author} {\bibinfo {author} {\bibfnamefont {N.}~\bibnamefont
  {Fl{\"a}schner}}, \bibinfo {author} {\bibfnamefont {B.~S.}\ \bibnamefont
  {Rem}}, \bibinfo {author} {\bibfnamefont {M.}~\bibnamefont {Tarnowski}},
  \bibinfo {author} {\bibfnamefont {D.}~\bibnamefont {Vogel}}, \bibinfo
  {author} {\bibfnamefont {D.-S.}\ \bibnamefont {Lühmann}}, \bibinfo {author}
  {\bibfnamefont {K.}~\bibnamefont {Sengstock}},\ and\ \bibinfo {author}
  {\bibfnamefont {C.}~\bibnamefont {Weitenberg}},\ }\bibfield  {title}
  {\bibinfo {title} {Experimental reconstruction of the berry curvature in a
  floquet bloch band},\ }\href {https://doi.org/10.1126/science.aad4568}
  {\bibfield  {journal} {\bibinfo  {journal} {Science}\ }\textbf {\bibinfo
  {volume} {352}},\ \bibinfo {pages} {1091} (\bibinfo {year}
  {2016})}\BibitemShut {NoStop}%
\bibitem [{\citenamefont {Gianfrate}\ \emph {et~al.}(2020)\citenamefont
  {Gianfrate}, \citenamefont {Bleu}, \citenamefont {Dominici}, \citenamefont
  {Ardizzone}, \citenamefont {{De Giorgi}}, \citenamefont {Ballarini},
  \citenamefont {Lerario}, \citenamefont {West}, \citenamefont {Pfeiffer},
  \citenamefont {Solnyshkov}, \citenamefont {Sanvitto},\ and\ \citenamefont
  {Malpuech}}]{Gianfrate2020}%
  \BibitemOpen
  \bibfield  {author} {\bibinfo {author} {\bibfnamefont {A.}~\bibnamefont
  {Gianfrate}}, \bibinfo {author} {\bibfnamefont {O.}~\bibnamefont {Bleu}},
  \bibinfo {author} {\bibfnamefont {L.}~\bibnamefont {Dominici}}, \bibinfo
  {author} {\bibfnamefont {V.}~\bibnamefont {Ardizzone}}, \bibinfo {author}
  {\bibfnamefont {M.}~\bibnamefont {{De Giorgi}}}, \bibinfo {author}
  {\bibfnamefont {D.}~\bibnamefont {Ballarini}}, \bibinfo {author}
  {\bibfnamefont {G.}~\bibnamefont {Lerario}}, \bibinfo {author} {\bibfnamefont
  {K.~W.}\ \bibnamefont {West}}, \bibinfo {author} {\bibfnamefont {L.~N.}\
  \bibnamefont {Pfeiffer}}, \bibinfo {author} {\bibfnamefont {D.~D.}\
  \bibnamefont {Solnyshkov}}, \bibinfo {author} {\bibfnamefont
  {D.}~\bibnamefont {Sanvitto}},\ and\ \bibinfo {author} {\bibfnamefont
  {G.}~\bibnamefont {Malpuech}},\ }\bibfield  {title} {\bibinfo {title}
  {{Measurement of the quantum geometric tensor and of the anomalous Hall
  drift}},\ }\href {https://doi.org/10.1038/s41586-020-1989-2} {\bibfield
  {journal} {\bibinfo  {journal} {Nature}\ }\textbf {\bibinfo {volume} {578}},\
  \bibinfo {pages} {381} (\bibinfo {year} {2020})}\BibitemShut {NoStop}%
\bibitem [{\citenamefont {Cuerda}\ \emph {et~al.}(2024)\citenamefont {Cuerda},
  \citenamefont {Taskinen}, \citenamefont {K\"allman}, \citenamefont
  {Grabitz},\ and\ \citenamefont {T\"orm\"a}}]{Cuerda2024}%
  \BibitemOpen
  \bibfield  {author} {\bibinfo {author} {\bibfnamefont {J.}~\bibnamefont
  {Cuerda}}, \bibinfo {author} {\bibfnamefont {J.~M.}\ \bibnamefont
  {Taskinen}}, \bibinfo {author} {\bibfnamefont {N.}~\bibnamefont {K\"allman}},
  \bibinfo {author} {\bibfnamefont {L.}~\bibnamefont {Grabitz}},\ and\ \bibinfo
  {author} {\bibfnamefont {P.}~\bibnamefont {T\"orm\"a}},\ }\bibfield  {title}
  {\bibinfo {title} {Observation of quantum metric and non-hermitian berry
  curvature in a plasmonic lattice},\ }\href
  {https://doi.org/10.1103/PhysRevResearch.6.L022020} {\bibfield  {journal}
  {\bibinfo  {journal} {Phys. Rev. Res.}\ }\textbf {\bibinfo {volume} {6}},\
  \bibinfo {pages} {L022020} (\bibinfo {year} {2024})}\BibitemShut {NoStop}%
\bibitem [{\citenamefont {Kim}\ \emph {et~al.}(2025)\citenamefont {Kim},
  \citenamefont {Chung}, \citenamefont {Qian}, \citenamefont {Park},
  \citenamefont {Jozwiak}, \citenamefont {Rotenberg}, \citenamefont {Bostwick},
  \citenamefont {Kim},\ and\ \citenamefont {Yang}}]{Kim2025}%
  \BibitemOpen
  \bibfield  {author} {\bibinfo {author} {\bibfnamefont {S.}~\bibnamefont
  {Kim}}, \bibinfo {author} {\bibfnamefont {Y.}~\bibnamefont {Chung}}, \bibinfo
  {author} {\bibfnamefont {Y.}~\bibnamefont {Qian}}, \bibinfo {author}
  {\bibfnamefont {S.}~\bibnamefont {Park}}, \bibinfo {author} {\bibfnamefont
  {C.}~\bibnamefont {Jozwiak}}, \bibinfo {author} {\bibfnamefont
  {E.}~\bibnamefont {Rotenberg}}, \bibinfo {author} {\bibfnamefont
  {A.}~\bibnamefont {Bostwick}}, \bibinfo {author} {\bibfnamefont {K.~S.}\
  \bibnamefont {Kim}},\ and\ \bibinfo {author} {\bibfnamefont {B.-J.}\
  \bibnamefont {Yang}},\ }\bibfield  {title} {\bibinfo {title} {Direct
  measurement of the quantum metric tensor in solids},\ }\href
  {https://doi.org/10.1126/science.ado6049} {\bibfield  {journal} {\bibinfo
  {journal} {Science}\ }\textbf {\bibinfo {volume} {388}},\ \bibinfo {pages}
  {1050} (\bibinfo {year} {2025})}\BibitemShut {NoStop}%
\bibitem [{\citenamefont {Guillot}\ \emph {et~al.}(2025)\citenamefont
  {Guillot}, \citenamefont {Blanchard}, \citenamefont {Pernet}, \citenamefont
  {Morassi}, \citenamefont {Lemaître}, \citenamefont {Gratiet}, \citenamefont
  {Harouri}, \citenamefont {Sagnes}, \citenamefont {Bloch},\ and\ \citenamefont
  {Ravets}}]{Guillot2025}%
  \BibitemOpen
  \bibfield  {author} {\bibinfo {author} {\bibfnamefont {M.}~\bibnamefont
  {Guillot}}, \bibinfo {author} {\bibfnamefont {C.}~\bibnamefont {Blanchard}},
  \bibinfo {author} {\bibfnamefont {N.}~\bibnamefont {Pernet}}, \bibinfo
  {author} {\bibfnamefont {M.}~\bibnamefont {Morassi}}, \bibinfo {author}
  {\bibfnamefont {A.}~\bibnamefont {Lemaître}}, \bibinfo {author}
  {\bibfnamefont {L.~L.}\ \bibnamefont {Gratiet}}, \bibinfo {author}
  {\bibfnamefont {A.}~\bibnamefont {Harouri}}, \bibinfo {author} {\bibfnamefont
  {I.}~\bibnamefont {Sagnes}}, \bibinfo {author} {\bibfnamefont
  {J.}~\bibnamefont {Bloch}},\ and\ \bibinfo {author} {\bibfnamefont
  {S.}~\bibnamefont {Ravets}},\ }\href {https://arxiv.org/abs/2507.16446}
  {\bibinfo {title} {A sublattice stokes polarimeter for bipartite photonic
  lattices}} (\bibinfo {year} {2025}),\ \Eprint
  {https://arxiv.org/abs/2507.16446} {arXiv:2507.16446 [cond-mat.mes-hall]}
  \BibitemShut {NoStop}%
\bibitem [{\citenamefont {Wunsch}\ \emph {et~al.}(2008)\citenamefont {Wunsch},
  \citenamefont {Guinea},\ and\ \citenamefont {Sols}}]{Wunsch2008}%
  \BibitemOpen
  \bibfield  {author} {\bibinfo {author} {\bibfnamefont {B.}~\bibnamefont
  {Wunsch}}, \bibinfo {author} {\bibfnamefont {F.}~\bibnamefont {Guinea}},\
  and\ \bibinfo {author} {\bibfnamefont {F.}~\bibnamefont {Sols}},\ }\bibfield
  {title} {\bibinfo {title} {Dirac-point engineering and topological phase
  transitions in honeycomb optical lattices},\ }\href
  {https://doi.org/10.1088/1367-2630/10/10/103027} {\bibfield  {journal}
  {\bibinfo  {journal} {New Journal of Physics}\ }\textbf {\bibinfo {volume}
  {10}},\ \bibinfo {pages} {103027} (\bibinfo {year} {2008})}\BibitemShut
  {NoStop}%
\bibitem [{\citenamefont {Montambaux}\ \emph {et~al.}(2009)\citenamefont
  {Montambaux}, \citenamefont {Pi\'echon}, \citenamefont {Fuchs},\ and\
  \citenamefont {Goerbig}}]{Montambaux2009}%
  \BibitemOpen
  \bibfield  {author} {\bibinfo {author} {\bibfnamefont {G.}~\bibnamefont
  {Montambaux}}, \bibinfo {author} {\bibfnamefont {F.}~\bibnamefont
  {Pi\'echon}}, \bibinfo {author} {\bibfnamefont {J.-N.}\ \bibnamefont
  {Fuchs}},\ and\ \bibinfo {author} {\bibfnamefont {M.~O.}\ \bibnamefont
  {Goerbig}},\ }\bibfield  {title} {\bibinfo {title} {Merging of dirac points
  in a two-dimensional crystal},\ }\href
  {https://doi.org/10.1103/PhysRevB.80.153412} {\bibfield  {journal} {\bibinfo
  {journal} {Phys. Rev. B}\ }\textbf {\bibinfo {volume} {80}},\ \bibinfo
  {pages} {153412} (\bibinfo {year} {2009})}\BibitemShut {NoStop}%
\bibitem [{\citenamefont {Montambaux}\ \emph {et~al.}(2018)\citenamefont
  {Montambaux}, \citenamefont {Lim}, \citenamefont {Fuchs},\ and\ \citenamefont
  {Pi\'echon}}]{Montambaux2018}%
  \BibitemOpen
  \bibfield  {author} {\bibinfo {author} {\bibfnamefont {G.}~\bibnamefont
  {Montambaux}}, \bibinfo {author} {\bibfnamefont {L.-K.}\ \bibnamefont {Lim}},
  \bibinfo {author} {\bibfnamefont {J.-N.}\ \bibnamefont {Fuchs}},\ and\
  \bibinfo {author} {\bibfnamefont {F.}~\bibnamefont {Pi\'echon}},\ }\bibfield
  {title} {\bibinfo {title} {Winding vector: How to annihilate two dirac points
  with the same charge},\ }\href
  {https://doi.org/10.1103/PhysRevLett.121.256402} {\bibfield  {journal}
  {\bibinfo  {journal} {Phys. Rev. Lett.}\ }\textbf {\bibinfo {volume} {121}},\
  \bibinfo {pages} {256402} (\bibinfo {year} {2018})}\BibitemShut {NoStop}%
\bibitem [{\citenamefont {Schneider}\ \emph {et~al.}(2016)\citenamefont
  {Schneider}, \citenamefont {Winkler}, \citenamefont {Fraser}, \citenamefont
  {Kamp}, \citenamefont {Yamamoto}, \citenamefont {Ostrovskaya},\ and\
  \citenamefont {Höfling}}]{Schneider2017}%
  \BibitemOpen
  \bibfield  {author} {\bibinfo {author} {\bibfnamefont {C.}~\bibnamefont
  {Schneider}}, \bibinfo {author} {\bibfnamefont {K.}~\bibnamefont {Winkler}},
  \bibinfo {author} {\bibfnamefont {M.~D.}\ \bibnamefont {Fraser}}, \bibinfo
  {author} {\bibfnamefont {M.}~\bibnamefont {Kamp}}, \bibinfo {author}
  {\bibfnamefont {Y.}~\bibnamefont {Yamamoto}}, \bibinfo {author}
  {\bibfnamefont {E.~A.}\ \bibnamefont {Ostrovskaya}},\ and\ \bibinfo {author}
  {\bibfnamefont {S.}~\bibnamefont {Höfling}},\ }\bibfield  {title} {\bibinfo
  {title} {Exciton-polariton trapping and potential landscape engineering},\
  }\href {https://doi.org/10.1088/0034-4885/80/1/016503} {\bibfield  {journal}
  {\bibinfo  {journal} {Reports on Progress in Physics}\ }\textbf {\bibinfo
  {volume} {80}},\ \bibinfo {pages} {016503} (\bibinfo {year}
  {2016})}\BibitemShut {NoStop}%
\bibitem [{\citenamefont {Shi}()}]{Shi2011}%
  \BibitemOpen
  \bibfield  {author} {\bibinfo {author} {\bibfnamefont {X.}~\bibnamefont
  {Shi}},\ }\bibfield  {title} {\bibinfo {title} {Joint {{Approximate
  Diagonalization Method}}},\ }in\ \href
  {https://doi.org/10.1007/978-3-642-11347-5_8} {\emph {\bibinfo {booktitle}
  {Blind {{Signal Processing}}: {{Theory}} and {{Practice}}}}},\ \bibinfo
  {editor} {edited by\ \bibinfo {editor} {\bibfnamefont {X.}~\bibnamefont
  {Shi}}}\ (\bibinfo  {publisher} {Springer})\ pp.\ \bibinfo {pages}
  {175--204}\BibitemShut {NoStop}%
\bibitem [{\citenamefont {Tarruell}\ \emph {et~al.}(2012)\citenamefont
  {Tarruell}, \citenamefont {Greif}, \citenamefont {Uehlinger}, \citenamefont
  {Jotzu},\ and\ \citenamefont {Esslinger}}]{Tarruell2012}%
  \BibitemOpen
  \bibfield  {author} {\bibinfo {author} {\bibfnamefont {L.}~\bibnamefont
  {Tarruell}}, \bibinfo {author} {\bibfnamefont {D.}~\bibnamefont {Greif}},
  \bibinfo {author} {\bibfnamefont {T.}~\bibnamefont {Uehlinger}}, \bibinfo
  {author} {\bibfnamefont {G.}~\bibnamefont {Jotzu}},\ and\ \bibinfo {author}
  {\bibfnamefont {T.}~\bibnamefont {Esslinger}},\ }\bibfield  {title} {\bibinfo
  {title} {{Creating, moving and merging Dirac points with a Fermi gas in a
  tunable honeycomb lattice}},\ }\href {https://doi.org/10.1038/nature10871}
  {\bibfield  {journal} {\bibinfo  {journal} {Nature}\ }\textbf {\bibinfo
  {volume} {483}},\ \bibinfo {pages} {302} (\bibinfo {year}
  {2012})}\BibitemShut {NoStop}%
\bibitem [{\citenamefont {Bellec}\ \emph {et~al.}(2013)\citenamefont {Bellec},
  \citenamefont {Kuhl}, \citenamefont {Montambaux},\ and\ \citenamefont
  {Mortessagne}}]{Bellec2013}%
  \BibitemOpen
  \bibfield  {author} {\bibinfo {author} {\bibfnamefont {M.}~\bibnamefont
  {Bellec}}, \bibinfo {author} {\bibfnamefont {U.}~\bibnamefont {Kuhl}},
  \bibinfo {author} {\bibfnamefont {G.}~\bibnamefont {Montambaux}},\ and\
  \bibinfo {author} {\bibfnamefont {F.}~\bibnamefont {Mortessagne}},\
  }\bibfield  {title} {\bibinfo {title} {Topological transition of dirac points
  in a microwave experiment},\ }\href
  {https://doi.org/10.1103/PhysRevLett.110.033902} {\bibfield  {journal}
  {\bibinfo  {journal} {Phys. Rev. Lett.}\ }\textbf {\bibinfo {volume} {110}},\
  \bibinfo {pages} {033902} (\bibinfo {year} {2013})}\BibitemShut {NoStop}%
\bibitem [{\citenamefont {Mili\ifmmode \acute{c}\else
  \'{c}\fi{}evi\ifmmode~\acute{c}\else \'{c}\fi{}}\ \emph
  {et~al.}(2019)\citenamefont {Mili\ifmmode \acute{c}\else
  \'{c}\fi{}evi\ifmmode~\acute{c}\else \'{c}\fi{}}, \citenamefont {Montambaux},
  \citenamefont {Ozawa}, \citenamefont {Jamadi}, \citenamefont {Real},
  \citenamefont {Sagnes}, \citenamefont {Lema\^{\i}tre}, \citenamefont
  {Le~Gratiet}, \citenamefont {Harouri}, \citenamefont {Bloch},\ and\
  \citenamefont {Amo}}]{Milicevic2019}%
  \BibitemOpen
  \bibfield  {author} {\bibinfo {author} {\bibfnamefont {M.}~\bibnamefont
  {Mili\ifmmode \acute{c}\else \'{c}\fi{}evi\ifmmode~\acute{c}\else
  \'{c}\fi{}}}, \bibinfo {author} {\bibfnamefont {G.}~\bibnamefont
  {Montambaux}}, \bibinfo {author} {\bibfnamefont {T.}~\bibnamefont {Ozawa}},
  \bibinfo {author} {\bibfnamefont {O.}~\bibnamefont {Jamadi}}, \bibinfo
  {author} {\bibfnamefont {B.}~\bibnamefont {Real}}, \bibinfo {author}
  {\bibfnamefont {I.}~\bibnamefont {Sagnes}}, \bibinfo {author} {\bibfnamefont
  {A.}~\bibnamefont {Lema\^{\i}tre}}, \bibinfo {author} {\bibfnamefont
  {L.}~\bibnamefont {Le~Gratiet}}, \bibinfo {author} {\bibfnamefont
  {A.}~\bibnamefont {Harouri}}, \bibinfo {author} {\bibfnamefont
  {J.}~\bibnamefont {Bloch}},\ and\ \bibinfo {author} {\bibfnamefont
  {A.}~\bibnamefont {Amo}},\ }\bibfield  {title} {\bibinfo {title} {Type-iii
  and tilted dirac cones emerging from flat bands in photonic orbital
  graphene},\ }\href {https://doi.org/10.1103/PhysRevX.9.031010} {\bibfield
  {journal} {\bibinfo  {journal} {Phys. Rev. X}\ }\textbf {\bibinfo {volume}
  {9}},\ \bibinfo {pages} {031010} (\bibinfo {year} {2019})}\BibitemShut
  {NoStop}%
\bibitem [{\citenamefont {Mondal}\ \emph {et~al.}(2024)\citenamefont {Mondal},
  \citenamefont {Ghadimi},\ and\ \citenamefont {Yang}}]{mondal2024b}%
  \BibitemOpen
  \bibfield  {author} {\bibinfo {author} {\bibfnamefont {C.}~\bibnamefont
  {Mondal}}, \bibinfo {author} {\bibfnamefont {R.}~\bibnamefont {Ghadimi}},\
  and\ \bibinfo {author} {\bibfnamefont {B.-J.}\ \bibnamefont {Yang}},\ }\href
  {https://arxiv.org/abs/2411.06724} {\bibinfo {title} {Non-abelian charge
  conversion in bilayer binary honeycomb lattice systems}} (\bibinfo {year}
  {2024}),\ \Eprint {https://arxiv.org/abs/2411.06724} {arXiv:2411.06724
  [cond-mat.mes-hall]} \BibitemShut {NoStop}%
\bibitem [{\citenamefont {Finck}\ \emph {et~al.}(2025)\citenamefont {Finck},
  \citenamefont {Solnyshkov}, \citenamefont {Dubois},\ and\ \citenamefont
  {Malpuech}}]{Finck2025}%
  \BibitemOpen
  \bibfield  {author} {\bibinfo {author} {\bibfnamefont {M.}~\bibnamefont
  {Finck}}, \bibinfo {author} {\bibfnamefont {D.}~\bibnamefont {Solnyshkov}},
  \bibinfo {author} {\bibfnamefont {J.}~\bibnamefont {Dubois}},\ and\ \bibinfo
  {author} {\bibfnamefont {G.}~\bibnamefont {Malpuech}},\ }\href
  {https://arxiv.org/abs/2507.19238} {\bibinfo {title} {Dirac points
  annihilation and its obstruction characterized by euler number and
  quaternionic charges in kagome lattice}} (\bibinfo {year} {2025}),\ \Eprint
  {https://arxiv.org/abs/2507.19238} {arXiv:2507.19238 [cond-mat.mes-hall]}
  \BibitemShut {NoStop}%
\bibitem [{\citenamefont {Nalitov}\ \emph {et~al.}(2015)\citenamefont
  {Nalitov}, \citenamefont {Solnyshkov},\ and\ \citenamefont
  {Malpuech}}]{Nalitov2015}%
  \BibitemOpen
  \bibfield  {author} {\bibinfo {author} {\bibfnamefont {A.~V.}\ \bibnamefont
  {Nalitov}}, \bibinfo {author} {\bibfnamefont {D.~D.}\ \bibnamefont
  {Solnyshkov}},\ and\ \bibinfo {author} {\bibfnamefont {G.}~\bibnamefont
  {Malpuech}},\ }\bibfield  {title} {\bibinfo {title} {Polariton $\mathbb{Z}$
  topological insulator},\ }\href
  {https://doi.org/10.1103/PhysRevLett.114.116401} {\bibfield  {journal}
  {\bibinfo  {journal} {Phys. Rev. Lett.}\ }\textbf {\bibinfo {volume} {114}},\
  \bibinfo {pages} {116401} (\bibinfo {year} {2015})}\BibitemShut {NoStop}%
\bibitem [{\citenamefont {Klembt}\ \emph {et~al.}(2018)\citenamefont {Klembt},
  \citenamefont {Harder}, \citenamefont {Egorov}, \citenamefont {Winkler},
  \citenamefont {Ge}, \citenamefont {Bandres}, \citenamefont {Emmerling},
  \citenamefont {Worschech}, \citenamefont {Liew}, \citenamefont {Segev},
  \citenamefont {Schneider},\ and\ \citenamefont {Höfling}}]{Klembt2018}%
  \BibitemOpen
  \bibfield  {author} {\bibinfo {author} {\bibfnamefont {S.}~\bibnamefont
  {Klembt}}, \bibinfo {author} {\bibfnamefont {T.~H.}\ \bibnamefont {Harder}},
  \bibinfo {author} {\bibfnamefont {O.~A.}\ \bibnamefont {Egorov}}, \bibinfo
  {author} {\bibfnamefont {K.}~\bibnamefont {Winkler}}, \bibinfo {author}
  {\bibfnamefont {R.}~\bibnamefont {Ge}}, \bibinfo {author} {\bibfnamefont
  {M.~A.}\ \bibnamefont {Bandres}}, \bibinfo {author} {\bibfnamefont
  {M.}~\bibnamefont {Emmerling}}, \bibinfo {author} {\bibfnamefont
  {L.}~\bibnamefont {Worschech}}, \bibinfo {author} {\bibfnamefont {T.~C.~H.}\
  \bibnamefont {Liew}}, \bibinfo {author} {\bibfnamefont {M.}~\bibnamefont
  {Segev}}, \bibinfo {author} {\bibfnamefont {C.}~\bibnamefont {Schneider}},\
  and\ \bibinfo {author} {\bibfnamefont {S.}~\bibnamefont {Höfling}},\
  }\bibfield  {title} {\bibinfo {title} {Exciton-polariton topological
  insulator},\ }\href {https://doi.org/10.1038/s41586-018-0601-5} {\bibfield
  {journal} {\bibinfo  {journal} {Nature}\ }\textbf {\bibinfo {volume} {562}},\
  \bibinfo {pages} {552} (\bibinfo {year} {2018})}\BibitemShut {NoStop}%
\bibitem [{\citenamefont {K\"onig}\ \emph {et~al.}(2023)\citenamefont
  {K\"onig}, \citenamefont {Yang}, \citenamefont {Budich},\ and\ \citenamefont
  {Bergholtz}}]{Konig2023}%
  \BibitemOpen
  \bibfield  {author} {\bibinfo {author} {\bibfnamefont {J.~L.~K.}\
  \bibnamefont {K\"onig}}, \bibinfo {author} {\bibfnamefont {K.}~\bibnamefont
  {Yang}}, \bibinfo {author} {\bibfnamefont {J.~C.}\ \bibnamefont {Budich}},\
  and\ \bibinfo {author} {\bibfnamefont {E.~J.}\ \bibnamefont {Bergholtz}},\
  }\bibfield  {title} {\bibinfo {title} {Braid-protected topological band
  structures with unpaired exceptional points},\ }\href
  {https://doi.org/10.1103/PhysRevResearch.5.L042010} {\bibfield  {journal}
  {\bibinfo  {journal} {Phys. Rev. Res.}\ }\textbf {\bibinfo {volume} {5}},\
  \bibinfo {pages} {L042010} (\bibinfo {year} {2023})}\BibitemShut {NoStop}%
\end{thebibliography}
\end{document}